\documentclass[sigconf, nonacm, pdfa]{acmart}
\usepackage{bbm,dsfont,subcaption,pdfpages,graphicx,placeins,makecell}
\usepackage{multirow} 
\usepackage[ruled,linesnumbered]{algorithm2e}
\usepackage{enumitem}
\newcommand\vldbdoi{10.14778/3717755.3717769}
\newcommand\vldbpages{1104 - 1117}
\newcommand\vldbvolume{18}
\newcommand\vldbissue{4}
\newcommand\vldbyear{2024}
\newcommand\vldbauthors{\authors}
\newcommand\vldbtitle{\shorttitle} 
\newcommand\vldbavailabilityurl{https://github.com/bytedance/adandv}
\newcommand\vldbpagestyle{empty}

\begin{document}

\title{\textsc{AdaNDV}: Adaptive Number of Distinct Value Estimation via Learning to Select and Fuse Estimators}



\author{Xianghong Xu}
\affiliation{
  \institution{ByteDance}
  \city{Beijing}
  \country{China}
}
\email{xuxianghong@bytedance.com}

\author{Tieying Zhang}
\affiliation{
  \institution{ByteDance}
  \city{San Jose}
  \country{USA}
}
\email{tieying.zhang@bytedance.com}\authornote{Tieying Zhang corresponds to this work.}

\author{Xiao He}
\affiliation{
  \institution{ByteDance}
  \city{Hangzhou}
  \country{China}
}
\email{xiao.hx@bytedance.com}

\author{Haoyang Li}
\affiliation{
  \institution{ByteDance}
  \city{Beijing}
  \country{China}
}
\email{lihaoyang.cs@bytedance.com}

\author{Rong Kang}
\affiliation{
  \institution{ByteDance}
  \city{Beijing}
  \country{China}
}
\email{kangrong.cn@bytedance.com}

\author{Shuai Wang}
\affiliation{
  \institution{ByteDance}
  \city{Beijing}
  \country{China}
}
\email{wangshuai.will@bytedance.com}

\author{Linhui Xu}
\affiliation{
  \institution{ByteDance}
  \city{Beijing}
  \country{China}
}
\email{xulinhui@bytedance.com}

\author{Zhimin Liang}
\affiliation{
  \institution{ByteDance}
  \city{Beijing}
  \country{China}
}
\email{liangzhimin@bytedance.com}

\author{Shangyu Luo}
\affiliation{
  \institution{ByteDance}
  \city{San Jose}
  \country{USA}
}
\email{shangyu.luo@bytedance.com}

\author{Lei Zhang}
\affiliation{
  \institution{ByteDance}
  \city{Shenzhen}
  \country{China}
}
\email{zhanglei.michael@bytedance.com}

\author{Jianjun Chen}
\affiliation{
  \institution{ByteDance}
  \city{San Jose}
  \country{USA}
}
\email{jianjun.chen@bytedance.com}

\begin{abstract}
Estimating the Number of Distinct Values (NDV) is fundamental for numerous data management tasks, especially within database applications.
However, most existing works primarily focus on introducing new statistical or learned estimators, while identifying the most suitable estimator for a given scenario remains largely unexplored. 
Therefore, we propose \textsc{AdaNDV}, a learned method designed to adaptively select and fuse existing estimators to address this issue. 
Specifically, (1) we propose to use learned models to distinguish between overestimated and underestimated estimators and then select appropriate estimators from each category. This strategy provides a complementary perspective by integrating overestimations and underestimations for error correction, thereby improving the accuracy of NDV estimation. 
(2) To further integrate the estimation results, we introduce a novel fusion approach that employs a learned model to predict the weights of the selected estimators and then applies a weighted sum to merge them. 
By combining these strategies, the proposed \textsc{AdaNDV} fundamentally distinguishes itself from previous works that directly estimate NDV.
Moreover, extensive experiments conducted on real-world datasets, with the number of individual columns being several orders of magnitude larger than in previous studies, demonstrate the superior performance of our method.
\end{abstract}

\maketitle

\pagestyle{\vldbpagestyle}
\begingroup\small\noindent\raggedright\textbf{PVLDB Reference Format:}\\
\vldbauthors. \vldbtitle. PVLDB, \vldbvolume(\vldbissue): \vldbpages, \vldbyear.\\
\href{https://doi.org/\vldbdoi}{doi:\vldbdoi}
\endgroup
\begingroup
\renewcommand\thefootnote{}\footnote{\noindent
This work is licensed under the Creative Commons BY-NC-ND 4.0 International License. Visit \url{https://creativecommons.org/licenses/by-nc-nd/4.0/} to view a copy of this license. For any use beyond those covered by this license, obtain permission by emailing \href{mailto:info@vldb.org}{info@vldb.org}. Copyright is held by the owner/author(s). Publication rights licensed to the VLDB Endowment. \\
\raggedright Proceedings of the VLDB Endowment, Vol. \vldbvolume, No. \vldbissue\ %
ISSN 2150-8097. \\
\href{https://doi.org/\vldbdoi}{doi:\vldbdoi} \\
}\addtocounter{footnote}{-1}\endgroup

\ifdefempty{\vldbavailabilityurl}{}{
\vspace{.3cm}
\begingroup\small\noindent\raggedright\textbf{PVLDB Artifact Availability:}\\
The source code, data, and/or other artifacts have been made available at \url{\vldbavailabilityurl}.
\endgroup
}

\section{Introduction}\label{sec:intro}
Identifying the Number of Distinct Values (NDV) in a data column is a fundamental task across numerous data management scenarios, particularly within the database domain. 
However, directly obtaining NDV is often impractical in real-world scenarios due to the prohibitive overheads of processing massive data volumes or data access restrictions.
Therefore, estimating NDV on limited samples has been a critical and longstanding research topic, explored for over seven decades~\cite{goodman1949estimation}. 
For instance, in the realm of Biology, a critical task is to estimate unseen species~\cite{valiant2013estimating,valiant2017estimating,mmo_bunge1993estimating}. 
Similarly, in Statistics, a recurring engagement involves quantifying the number of distinct categories within a given population~\cite{goodman1949estimation,chao1984nonparametric}.
Moreover, in the domain of Networks, assessing the quantity of virtualized devices is a significant challenge~\cite{network_cohen2019cardinality,network_nath2008synopsis}. 

In Database, some widely used systems (e.g., Spark and PostgreSQL) directly rely on NDV to compute cardinality, a metric that is subsequently utilized by the query optimizer~\cite{spark_plan_code,pg_plan_code}. 
Besides, NDV affects the join order selection in MySQL as well~\cite{mysql_join}.
Furthermore, recent studies show that precise NDV estimation can generate better query plans that bring significant SQL query execution latency reductions~\cite{li2023alece,han2024bytecard}.

\begin{figure}
    \centering
    \includegraphics[width=0.8\linewidth]{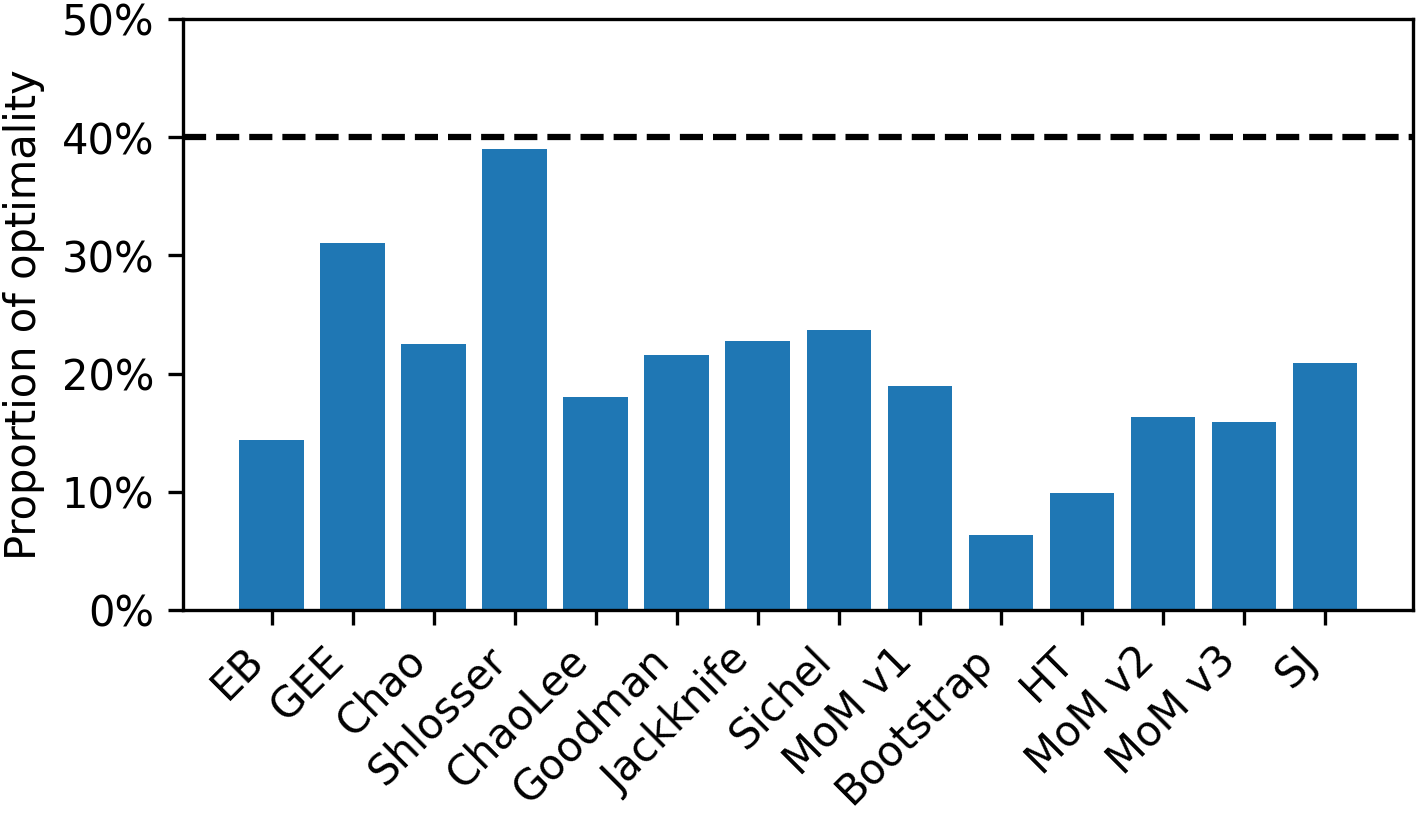}
    \caption{Evaluation of fourteen statistical estimators on 25,159 test columns, where the bar represents the proportion of each estimator achieving the optimality (lowest estimation error) among the fourteen estimators. No single estimator achieves optimality on more than 40\% of test cases.}
    \label{fig:traditional-strength}
\end{figure}

Prolonged research efforts in NDV estimation have accumulated numerous NDV estimators, the majority of which are statistical methods relying on manually designed polynomials computing or equation-solving~\cite{goodman1949estimation,gee_charikar2000towards,error_bound,chao_in_db_ozsoyoglu1991estimating,chao1984nonparametric,shlosser1981estimation,chaolee,hybskew_haas1995sampling,sichel1986parameter,sichel1986word,sichel1992anatomy,mmo_bunge1993estimating,bootstrap_smith1984nonparametric,horvitz_sarndal1992model,hybskew_haas1995sampling}.
Each statistical estimator is grounded in distinct hypotheses regarding the underlying distribution. As a result, their efficacy markedly declines when the actual distribution does not conform to their assumptions.
Recently, some studies have introduced learned estimators based on Machine Learning (ML) techniques~\cite{li2024learning,ls_wu2022learning} to solve this issue, demonstrating better performance than statistical estimators.
Nevertheless, despite the significant progress achieved, the domain of NDV estimation is still beset by the following difficulties:

\noindent\textbf{(1) Selection dilemma.} 
Although a plethora of estimators exist, there frequently remains ambiguity regarding the optimal choice for practical application.
The question of \textit{which estimator is most suitable for a specific data column} has received scarce attention over time. 
As new estimators continue to be introduced, this understudied question becomes increasingly substantial. 
To intuitively show the selection dilemma, we depict the performance of fourteen statistical estimators on a large dataset comprising 25,159 test cases (for additional details, please refer to Section~\ref{sec:exp-settings}) in Figure \ref{fig:traditional-strength}.
The results illustrate that no single estimator consistently achieves the lowest estimation error across all test cases, {indicating that no individual estimators can consistently beat others}. For instance, the top-performing estimator, Shlosser, only manages to outperform others in roughly 40\% of the cases.  
This observation highlights the intricate nature of selecting a suitable estimator from a set of available options.

\noindent\textbf{(2) Underexploitation issue.}
Most studies have focused on exploring new estimators, including the recently proposed learned estimator~\cite{li2024learning,ls_wu2022learning}, while exploiting existing estimators to improve estimations has been largely overlooked. 
To better illustrate the issue, we conceptualize a \textit{hypothetical} estimator, which involves picking one of the aforementioned fourteen statistical estimators, under the hypothetical condition that we know the actual estimation errors in advance. 
Precisely, we name the hypothetical estimator ``Hypo-optimal'' since we select the estimator with the lowest actual estimation error for every test.
The performance of the hypothetical estimator and a state-of-the-art (SOTA) learned estimator~\cite{ls_wu2022learning} is shown in Table~\ref{tab:ideal-traditional}. 
From the results in the table, it is evident that the Hypo-optimal estimator considerably outperforms the SOTA learned estimator. 
The comparison illustrates the substantial potential held by statistical estimators, simultaneously underscoring the critical significance of judiciously exploiting estimators.

\begin{table}[t]
    \centering
    \caption{Experiments on hypothetical estimators. The numbers represent estimation errors, with lower values indicating better performance.}
    \begin{tabular}{ccccccc}
\toprule
        Estimator& Mean & 50\% & 75\%& 90\% & 95\% & 99\% \\
\midrule
         Hypo-optimal  & \textbf{1.20} & \textbf{1.08} & \textbf{1.28} & \textbf{1.56} & \textbf{1.83} & \textbf{2.36} \\
        \makecell{\footnotesize SOTA learned estimator \\ \footnotesize ~\cite{ls_wu2022learning}}  & 2.24 & 1.72 & 2.28 & 3.20 & 4.11 & 10.46 \\
\bottomrule
    \end{tabular}
    \label{tab:ideal-traditional}
\end{table}

Inspired by the above observations, in this paper, we introduce \textsc{AdaNDV}, an \underline{Ada}ptive \underline{NDV} estimation method learning to select the proper estimators from existing ones and to fuse their estimation results to enhance estimation precision for different scenarios.
Selecting the optimal estimator with high accuracy is quite challenging because it is difficult to extract adequate features and explore appropriate ML models.
{On the contrary, selecting the $k$ estimators that are most likely to approach the ground truth is a relatively easier task for ML models~\cite{liu2009learning}. 
}
Moreover, we distinctively propose to address the issue by distinguishing whether an estimator is overestimated or underestimated, allowing us to utilize them accordingly.
{Specifically, if one estimator overestimates and another underestimates for a test case, there exists a set of weights such that their weighted sum performs better than either estimator individually.}
This approach inherently leverages the complementary nature of overestimations and underestimations to reduce estimation error. 
{In addition, in our initial experiments, we found that certain base estimators tend to make more overestimations or underestimations, indicating that distinguishing between overestimating and underestimating estimators is a comparatively easier task.}
Subsequently, we select the estimators with leading overestimation and underestimation performance respectively.
Further, we introduce a learned model to predict the weights of the chosen estimators and then establish the ultimate estimation by applying a weighted sum to fuse them. 
Different from the previous works that directly estimate NDV, our method offers a novel approach to enhance the accuracy of NDV estimation by merging existing estimators.
This allows our method to adaptively select appropriate estimators for specific scenarios and allocate proper weights to fuse them for end-to-end estimation.
Finally, extensive experiments are carried out on a voluminous dataset from the real world. Specifically, the number of individual test columns exceeds tens of thousands, while previous works were tested on at most about two hundred columns. This orders of magnitude increase in individual test columns enables a thorough evaluation encompassing existing estimators alongside our novel approach.

To sum up, the main contributions are shown as follows:
\begin{itemize}
    \item We propose \textsc{AdaNDV}, an adaptive NDV estimation method learning to select and fuse appropriate estimators for specific scenarios. To the best of our knowledge, we are the first to combine existing estimators with ML techniques to improve NDV estimation. 
    \item We introduce an innovative overestimation-underestimation complementary perspective for estimator selection and exploitation to reduce estimation error.
    \item We develop a novel learned weighted sum strategy to fuse the estimation results to obtain the ultimate estimation, which is significantly different from directly estimating NDV.
    \item Extensive experiments, conducted on a rich, real-life dataset with tens of thousands of individual columns, significantly larger than at most hundreds of columns used in past studies, demonstrate the superiority of \textsc{AdaNDV}.
\end{itemize}

\begin{figure*}[t]
    \centering
    \includegraphics[width=0.9\linewidth]{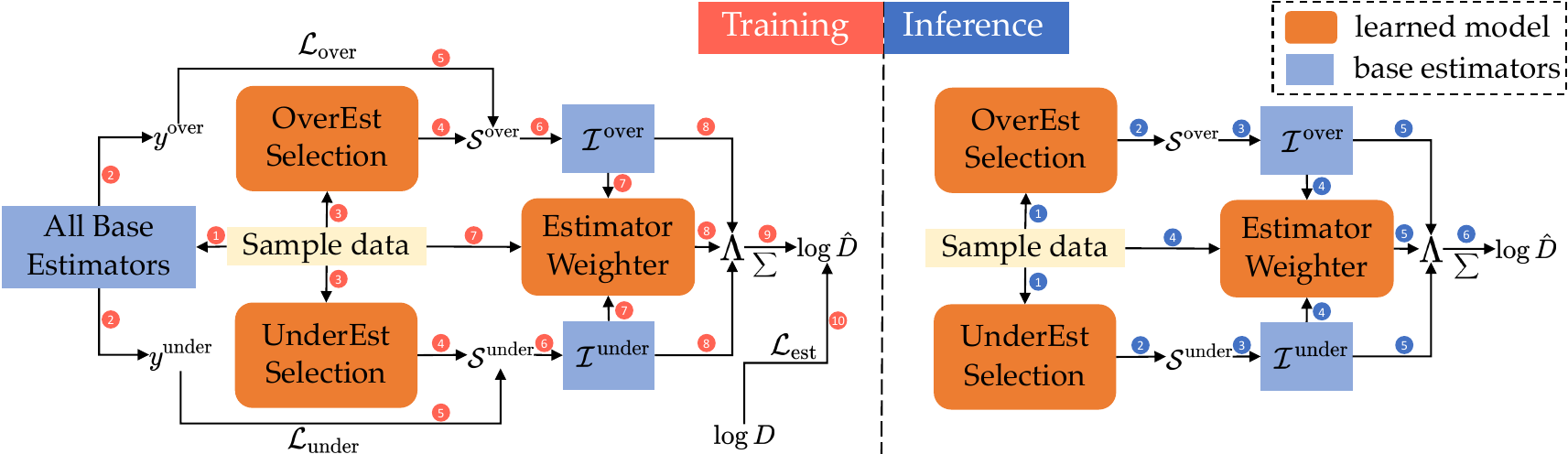}
    \caption{Overview of \textsc{AdaNDV} on NDV estimation including training and inference data pipelines.}
    \label{fig:model_pipeline}
\end{figure*}

\section{Preliminaries}
\subsection{Problem Statement}\label{sec:statement}
Existing NDV estimation methods can be categorized into \textit{sketch-based}~\cite{harmouch2017cardinality,flajolet2007hyperloglog,ertl2023ultraloglog} and \textit{sampling-based}~\cite{goodman1949estimation,ls_wu2022learning,li2024learning}, while most of the former require scanning all the data, making them impractical in many scenarios. Here, we focus on sampling-based NDV estimation.

\noindent\textbf{NDV Estimation Definition.} Given a data column $C$ with $N$ rows, let $D$ be the NDV of column $C$. The task is to estimate $D$ by uniformly sampling $n$ $(n\leq N)$ rows (denoted $S$, and $S\subseteq  C$) from $C$. Let $r=n/N$ be the sampling rate, and we assume $N$ is known. We define $d$ as the NDV of $S$. 

Two features are widely discussed in sampling-based NDV estimation: \textit{frequency} and \textit{frequency profile}.

\noindent\textbf{Frequency.} The frequency of a value $x$ in $C$ is the number of times it appears in $C$. Denote $N_x=\sum_{i\in C} \mathds{1}_x(i)$, where $N_x$ is the frequency of value $x$ in $C$, $\mathds{1}_x(\cdot)$ is the indicator function that returns 1 if the input equals to $x$ and 0 otherwise. Similarly, $n_x=\sum_{i\in S} \mathds{1}_x(i)$ is the frequency of value $x$ in $S$.

\noindent\textbf{Frequency Profile.} Frequency profile is the \textit{frequency of frequency}. Let the frequency profile of $C$ be $F=(F_j)_{j=1,2,\ldots,N}$, where $F_j=|\{i\in C|N_i=j\}|$. Similarly, the frequency profile of $S$ be $f=(f_j)_{j=1,2,\ldots,n}$, where $f_j=|\{i\in S|n_i=j\}|$.

\noindent\textbf{Feature Relations and Examples.} The two features are closely related to NDV and the number of rows. We can get $D=|\{i\in C|N_i>0\}|=\sum_{j=1}^N F_j$ and $d=|\{i\in S|n_i>0\}|=\sum_{j=1}^n f_j$. Besides, the total number of rows can be expressed as $N=\sum_{j=1}^N j\cdot F_j$, and $n=\sum_{j=1}^n j\cdot f_j$. 

For instance, suppose the sample data is $S=\{a, a, a, b, b, b, c, c, d\}$, we can observe that the sample size is $|S|=9$, and $d=4$ (there are $a,b,c,d$ four distinct values). The frequency of $S$ is $\{n_a=3,n_b=3,n_c=2,n_d=1\}$, and the frequency of frequency is $\{f_1=1,f_2=1,f_3=2,f_i=0,i=4,\ldots,9\}$. Based on these features, we can get $d=\sum_{i=1}^9f_i=4$, and $|S|=\sum_{i=1}^9i\cdot f_i=9$.

\subsection{NDV Estimators}
We use several representative estimators to demonstrate how they use the frequency profile of sample data to estimate NDV.

\noindent\textbf{Traditional Estimators.}
Goodman~\cite{goodman1949estimation} is a representative \textit{linear polynomial} estimator with a sophisticated expression:
\begin{equation}
D_{\mathrm{Goodman}} =d+\sum_{i=1}^n(-1)^{i+1} \frac{(N-n+i-1) !(n-i) !}{(N-n-1) ! n !} f_i.
    \label{eq:goodman}
\end{equation}

Chao~\cite{chao1984nonparametric,chao_in_db_ozsoyoglu1991estimating} estimator has a \textit{nonlinear polynomial} expresion:
\begin{equation}
    D_{\mathrm{Chao}} = d + \frac{f_1^2}{2f_2}.
    \label{eq:chao}
\end{equation}

Besides, some estimators need to solve sophisticate \textit{non-linear equations} constructed by the frequency profiles~\cite{gee_charikar2000towards,sichel1986parameter,sichel1986word,sichel1992anatomy,mmo_bunge1993estimating}. For instance, Sichel~\cite{sichel1986parameter,sichel1986word,sichel1992anatomy} estimator needs to solve the following non-linear equations:

\begin{align}
    \begin{aligned}
        (1+g)\ln g-Ag+B=0,\frac{f_1}{n}<g<1, A=\frac{2n}{d}-\ln\frac{n}{f_1},\\
        B=\frac{2f_1}{d}+\ln \frac{n}{f_1},\hat{b}=\frac{g\ln \frac{ng}{f_1}}{1-g},\hat{c}=\frac{1-g^2}{ng^2},
        D_{\mathrm{Sichel}}=\frac{2}{\hat{b}\hat{c}}.\\
    \end{aligned}
    \label{eq:sichel}
\end{align}

\noindent\textbf{Learned Estimators.}
Recently, ML techniques have been introduced into NDV estimation~\cite{ls_wu2022learning,li2024learning}. Wu et al.~\cite{ls_wu2022learning} proposed a learned statistician (LS in short) to estimate NDV. It constructs a multi-layer perception (MLP) as the estimator, which takes the cut-off of the frequency profile and some features as input and outputs the estimated NDV. It is trained in the regression paradigm, which minimizes the $L_2$ loss between the estimated NDV and the ground truth.

\noindent\textbf{Evaluation Protocol.} 
Ratio-error, also known as q-error~\cite{q_error_moerkotte2009preventing}, is widely used to evaluate the performance of an estimator in database applications:
\begin{equation}
    \mathrm{q\text{-}error}=\max(\frac{\hat{D}}{D},\frac{D}{\hat{D}}),
    \label{eq:q-error}
\end{equation}
where $\hat{D}$ is the estimated NDV and $D$ is the ground truth NDV. The lower error represents the better performance.

\section{Methodology}\label{sec:methodology}
\subsection{Model Architecture}\label{sec:architecture}

\subsubsection{Overview of \textsc{AdaNDV}}\label{sec:overview-base}
The architecture of \textsc{AdaNDV} is shown in Figure~\ref{fig:model_pipeline} and there are four components.

\noindent (1) Base estimators collection. We collect fourteen representative statistical estimators as our base estimators, detailed in Section \ref{sec:exp-settings}. We do not include learned estimators due to the limited availability of such estimators and nontrivial overhead associated with training them. Additionally, statistical estimators offer higher efficiency and are free of training.

\noindent (2) Leading estimator selection. ``OverEst'' and ``UnderEst'' represent overestimation and underestimation. Estimator selection is designed to prioritize the base estimators by their overestimation and underestimation errors. Essentially, the prioritization of a set of base estimators entails assigning them scores that reflect their performance in terms of overestimation and underestimation errors. 
Specifically, we use two identical learned models with different training objectives to prioritize the two types of estimators. The loss functions of the two models are denoted as $\mathcal{L}_{\mathrm{over}}$ and $\mathcal{L}_{\mathrm{under}}$, respectively. This component will be detailed in Section~\ref{sec:leading}.

\noindent (3) Estimator fusion. We first select estimators with the top-$k$ overestimated and underestimated performance, denoted as $\mathcal{I}^{\mathrm{over}}$ and $\mathcal{I}^{\mathrm{under}}$. Next, we use a learned model to assign weights to each selected estimator, and then employ a weighted sum to compute the NDV. 
For instance, suppose $k$ is 1, the ground truth NDV $D$ is 10,000, and the estimation results of the selected estimators are $\hat{D}_1=11,000$ and $\hat{D}_2=9,000$, our estimation is formulated as $\hat{D}=\Lambda_1\hat{D}_1+\Lambda_2\hat{D}_2, 0\leq \Lambda_1,\Lambda_2\leq 1, \Lambda_1+\Lambda_2=1$. This fusion method can reduce estimation errors based on existing estimators.
The loss function of this component is denoted as $\mathcal{L}_{\mathrm{est}}$ and its details will be elaborated in Section~\ref{sec:exploit}.

\noindent (4) Model training. There are three objectives in \textsc{AdaNDV}, and we can derive the end-to-end loss function to train our method:

\begin{align}
    \mathcal{L}_{\mathrm{\textsc{AdaNDV}}}=\mathcal{L}_{\mathrm{over}}+\mathcal{L}_{\mathrm{under}}+\beta \mathcal{L}_{\mathrm{est}},
    \label{eq:loss}
\end{align}
where $\beta$ is a hyperparameter that modulates the trade-offs between different kinds of training objectives. Our proposed method can be trained by minimizing the $\mathcal{L}_{\mathrm{\textsc{AdaNDV}}}$ loss function.

\noindent\textbf{Training pipeline.}
\textcircled{1}All base estimators use the sample data to estimate NDV. \textcircled{2}It is straightforward to distinguish the overestimated and underestimated estimators on the training data, and then we construct training \underline{labels} $y^{\mathrm{over}}$ and $y^{\mathrm{under}}$. \textcircled{3}-\textcircled{4}The estimator selection models will respectively generate the \underline{scores} that prioritize the estimators by the predicted overestimation and underestimation performance. $\mathcal{S}^{\mathrm{over}}\in\mathbb{R}^m$ represents the scores based on overestimation, where $m$ is the number of base estimators, with a higher value indicating better overestimation performance. $\mathcal{S}^{\mathrm{under}}\in\mathbb{R}^m$ is similar to $\mathcal{S}^{\mathrm{over}}$.
\textcircled{5}The \underline{selection loss functions} ($\mathcal{L}_{\mathrm{over}}$ and $\mathcal{L}_{\mathrm{under}}$) of the two models take the scores and labels as input. \textcircled{6}Then we respectively select top base estimators with high scores and the \underline{selected estimators} are $\mathcal{I}^{\mathrm{over}}$ and $\mathcal{I}^{\mathrm{under}}$. \textcircled{7}The learned estimator weighter takes the sample data and the estimations of the selected estimators as input and predicts the \underline{weight} ($\Lambda$). \textcircled{8}-\textcircled{9}Finally, we employ a weighted sum on the estimation results of the selected estimators to fuse them into the \underline{ultimate NDV estimation} $\hat{D}$, \textcircled{\footnotesize 10}deriving \underline{fusion-based estimation loss function} $\mathcal{L}_{\mathrm{est}}$.

\noindent\textbf{Inference pipeline.} \textcircled{1}-\textcircled{2}The learned leading estimator selection models take the sample data as input to generate the scores of each estimator. \textcircled{3}Then, we select the top estimators with the highest scores for overestimation and underestimation, respectively. \textcircled{4}Next, we input the sample data and the estimations of the selected estimator into the learned estimator weighter to obtain the weights. \textcircled{5}-\textcircled{6}Finally, the ultimate estimation is fused by a weighted sum on the estimations of the selected estimators.

\begin{figure}[t]
    \centering
    \includegraphics[width=0.96\linewidth]{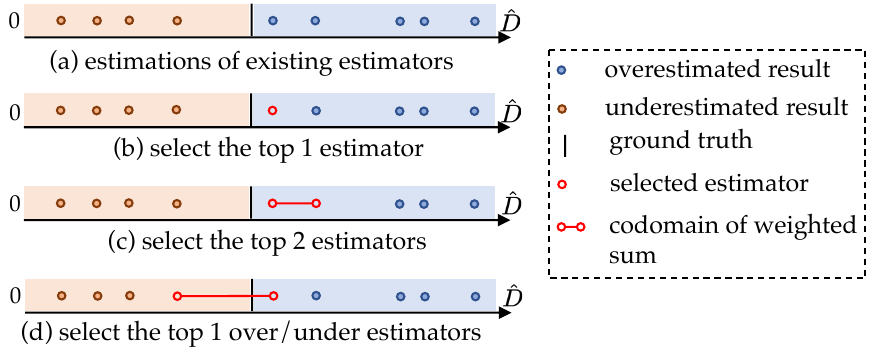}
    \caption{Intuition behind leveraging the properties of overestimation and underestimation.}
    \label{fig:properties}
\end{figure}

\subsubsection{Properties of overestimation and underestimation}\label{sec:motivation}
We illustrate the intuition behind our method in leveraging the properties of overestimation and underestimation in Figure~\ref{fig:properties}. Each circle in the figure represents the estimation result from a base estimator. The ground truth, indicated by a vertical bar in the middle, separates these results into two sections: overestimation and underestimation, which are marked with two different colors.
The performance of selecting the optimal estimator from existing ones (as shown in Figure~\ref{fig:properties}b) is challenging and limited by both selection accuracy and the base estimators. 
While it is straightforward to ensemble estimators to alleviate these issues, selecting estimators with the lowest errors and ensemble them may not bring performance improvement. Figure~\ref{fig:properties}c illustrates such a scenario where two overestimation results are selected, and the corresponding codomain cannot cover the ground truth bar.
Therefore, differentiating the results into overestimations and underestimations is essential for improving the performance. Although the estimation errors of the selected results may be substantial, complementing overestimations and underestimations enables their weighted sum to robustly encompass the ground truth, as shown in Figure~\ref{fig:properties}d. This allows the model to learn a set of parameters, resulting in the weighted sum of the selected results performing better than any individual estimator, potentially even approaching the ground truth.

\subsection{Leading Estimator Selection}\label{sec:leading}
We first describe the features extracted from the sample data, then we comprehensively illustrate the ranking paradigm used in estimator selection, and finally, we delineate the construction of the objective function for this component.

\subsubsection{Feature Engineering} \label{sec:fe}
Formally, we denote the features extracted from the sample data as $x\in\mathbb{R}^{ H}$, where $H$ denotes the number of dimensions in the feature space. Frequency profiles $f$ are widely used features, but their sizes varies across different column test cases. Since a learned model needs a fixed number of input features, we apply a cut-off to the frequency profile, similar to previous works~\cite{ls_wu2022learning,li2024learning}. This operation is based on the assumption that the predictive power of $f_i$ decreases as $i$ increases, an assumption that is widely used in previous works~\cite{ls_wu2022learning,li2024learning,chao1984nonparametric,gee_charikar2000towards,hybskew_haas1995sampling}. In addition, according to Section \ref{sec:statement}, cutting off the frequency profile will make computing the number of sample data $n$ and the NDV of sample data $d$ inapplicable. Therefore, we use $n$ and $d$ as directly as features. Furthermore, we include the original column size $N$ as another feature.

Since the length of input features must be a fixed number $H$, if the size of $f$ is smaller than the required length, we pad it with zeros. Specifically, the input feature is formulated as:

\begin{align}
    x = [f_1, f_2, \cdots, f_{H-3}, \log n, \log d, \log N].
\end{align}
We apply the logarithm operation on $n$, $d$, and $N$ to mitigate the skewness of input features.

\subsubsection{Estimator Ranking} 
The ranking paradigm has demonstrated substantial superiority in solving item prioritization tasks~\cite{liu2009learning,bruch2019revisiting,listmle_xia2008listwise,TensorflowRankingKDD2019,wang2018lambdaloss}. We adapt SOTA ranking techniques in estimator selection, which encompasses the elaborated construction of ranking labels and the training of the learned ranker.

\noindent\textbf{Ranking Label Construction.} Constructing the ranking label is a complex task~\cite{wang2013theoretical} due to the potentially infinite number of label values, even when the ranking order is fixed. For instance, given two estimators $e_1$ and $e_2$, where $e_1$ has a lower q-error. We denote $e_1\succ e_2$, indicating that $e_1$ is better than $e_2$. Let $y_1$ and $y_2$ be the ranking labels. Any values of $y_1$ and $y_2$ that satisfy the condition $y_1>y_2$ are eligible to serve as ranking labels, because the ranking paradigm focuses solely on the relative orders rather than specific values. 

There are no predefined estimator ranking labels available, and the labels in the estimator selection context require the following attributes: (1) A \textit{higher} value of the label reflects the \textit{higher} priority, while also indicating a \textit{lower} q-error in NDV estimation; (2) The value domain of the label needs to be constrained, as applying ranking techniques usually involves exponentiation operation on the label values~\cite{liu2009learning}. Unconstrained values could potentially lead to computational overflow; (3) The label needs to distinctly differentiate between overestimated and underestimated estimators, as required by our designed objectives. 

To this end, we propose an efficient ranking label construction strategy. Firstly, we use the ranking position to constrain the value of labels to be no greater than $m$. Then, we construct the labels by reversing the ranking positions based on the lowest overestimated or underestimated q-error. Finally, we mask the overestimated or underestimated estimators to differentiate them. Through these steps, the constructed labels satisfy the required attributes.

\begin{algorithm}[t]
\SetAlgoLined
\KwIn{$\hat{\mathcal{D}},D$}
\KwOut{$y^{\mathrm{over}}$}
 \textit{UnderEstSet} $\xleftarrow{} \emptyset$; $i \xleftarrow{} 1$; $\hat{\mathcal{D}}_{\mathrm{max}}\xleftarrow{}\max_{1\leq j \leq m}\hat{\mathcal{D}}_j$;\\
 \For{$i\xleftarrow{}1;i\leq m;i\xleftarrow{}i+1$}{
    \If{$\hat{\mathcal{D}}_i\leq D$}{
   \textit{UnderEstSet} $\xleftarrow{} i$ \; 
   $\hat{\mathcal{D}}_i\xleftarrow{}\hat{\mathcal{D}}_i + \hat{\mathcal{D}}_{\mathrm{max}} $; \\
   }
 }
\For{$i\xleftarrow{}1;i\leq m;i\xleftarrow{}i+1$} {
$y^{\mathrm{over}}_i\xleftarrow{}m-\pi_{\hat{\mathcal{D}}_i}$;   \\
}
\For{$i$ in UnderEstSet} {
$y^{\mathrm{over}}_i\xleftarrow{} 0$; // mask the underestimate estimators
}
 \textbf{return} $y^{\mathrm{over}}$\;
 \caption{Overestimation ranking label construction.}\label{algo:over_label}
\end{algorithm}

To prioritize the estimators with low overestimation q-errors, the process of constructing ranking labels $y^{\mathrm{over}}$ is shown in Algorithm~\ref{algo:over_label}. Specifically, we first record the maximum estimated result among all base estimators as $\hat{\mathcal{D}}_{\mathrm{max}}$. Then, to separate underestimations from overestimations, we add $\hat{\mathcal{D}}_{\mathrm{max}}$ to the result of each underestimated estimator to ensure that their estimations are higher than those of any overestimated estimators. Next, we obtain the ranking position $\pi$ of each base estimator by sorting the estimation results, where $\pi_{\hat{\mathcal{D}}}=\operatorname{argsort}(\hat{\mathcal{D}})$. Specifically, the $\operatorname{argsort}$ operation yields the indices required to sort the data in ascending order, and $\pi_{\hat{\mathcal{D}_i}}$ represents the index of $\hat{\mathcal{D}_i}$. In the context of overestimation, the transformed estimation is considered better when it is smaller. Therefore, the estimator with better overestimation performance has a higher value of the ranking label $y_i^{\mathrm{over}}$ by reversing its ranking position. Finally, we mask the underestimated estimators by setting their ranking labels as zero to differentiate them from the overestimated ones. 

The process of constructing labels $y^{under}$ is similar to that of $y^{over}$, and we omit it for conciseness.

\noindent\textbf{Train the Learned Ranker.} 
We use a multi-layer perceptron (MLP) as the backbone of our ranking model: $\mathcal{S}=\mathrm{MLP}(x)$, where $\mathcal{S}\in\mathbb{R}^m$ represents the ranking scores of the estimators. The higher score indicates the higher priority of the corresponding estimator. We adapt the SOTA ranking techniques~\cite{bruch2019revisiting,listmle_xia2008listwise,TensorflowRankingKDD2019,wang2018lambdaloss} into estimator selection to train the ranking models. In short, the learned ranker can be trained by:

\begin{align}
\begin{aligned}
    \mathcal{L}_{\mathrm{rank}}(\mathcal{S},y)&= - \sum_{i=1}^m\frac{2^{y_i}-1}{\log_2(1+\pi(i))} ,\\
    \pi(i)&=1+\sum_{j,j\ne i}^m\frac{1}{1+e^{-\alpha(\mathcal{S}_j-\mathcal{S}_i)}},
\end{aligned}
    \label{eq:ranking}
\end{align}
where $\alpha$ is the hyperparameter in the training framework, $\mathcal{L}_{\mathrm{rank}}$ is the estimator ranking loss function, and $y$ is the ranking label. The learned ranker can similarly achieve different objectives by assigning the labels constructed by Algorithm~\ref{algo:over_label}.

\subsubsection{Leading Estimator Selection} In this subsection, we show the training objectives of complementary estimator selection of this component.

\noindent\textbf{Overestimated Estimators Selection.}
We construct the training labels $y^{\mathrm{over}}$ according to Algorithm~\ref{algo:over_label}, and we use an MLP that has two hidden layers with 128 and 64 dimensions to compute the ranking scores $\mathcal{S}^{\mathrm{over}}$. The loss function for overestimated estimator selection is:

\begin{align}
    \mathcal{L}_{\mathrm{over}}=\frac{1}{\mathcal{N}}\sum_{i=1}^\mathcal{N}\mathcal{L}_{\mathrm{rank}}({\mathcal{S}_i^{\mathrm{over}}},y^{\mathrm{over}}_i),
    \label{eq:loss-over}
\end{align}
where $\mathcal{N}$ is the number of training samples, and $\mathcal{L}_{\mathrm{rank}}$ is the loss function defined in Equation (\ref{eq:ranking}).

\noindent\textbf{Underestimated Estimators Selection.} Similarly, we use another MLP with the identical model architecture to compute $\mathcal{S}^{\mathrm{under}}$ and the training labels $y^{\mathrm{under}}$.  The loss function for underestimated estimator selection is:

\begin{align}
    \mathcal{L}_{\mathrm{under}}=\frac{1}{\mathcal{N}}\sum_{i=1}^\mathcal{N}\mathcal{L}_{\mathrm{rank}}({\mathcal{S}_i^{\mathrm{under}}},y^{\mathrm{under}}_i).
    \label{eq:loss-under}
\end{align}

\subsection{Estimator Fusion}\label{sec:exploit}
\subsubsection{Feature Engineering} The features described in Section~\ref{sec:fe} are also used in this component. Moreover, we incorporate the estimated results of the chosen base estimators as the additional features. 

Specifically, the leading estimator selection component provides the priority scores $\mathcal{S}^{\mathrm{over}}$ and $\mathcal{S}^{\mathrm{under}}$of the base estimators. We select $k$ estimators with the top-$k$ highest scores for both both overestimated and underestimated estimators. The chosen estimators are denoted as $\mathcal{I}^{\mathrm{over}}=\operatorname{argmax}_k\mathcal{S}^{\mathrm{over}}$ and $\mathcal{I}^{\mathrm{under}}=\operatorname{argmax}_k\mathcal{S}^{\mathrm{under}}$. The features are defined as $x^\prime=[x,\hat{\mathcal{D}}|_{\mathcal{I}^{\mathrm{over}}},\hat{\mathcal{D}}|_{\mathcal{I}^{\mathrm{under}}}],x^\prime\in\mathbb{R}^{H+2k}$.

\subsubsection{Estimator Fusion} \label{sec:fusion}
We use an MLP to compute the weights for the chosen base estimators: $\Lambda=\mathrm{MLP}(x^\prime)$, where $\Lambda\in\mathbb{R}^{2k}$ is the weight vector concatenated by two $k$-dimentional vectors corresponding to the chosen leading estimators. $\Lambda$ is not fixed but depends on sample data and selected estimators. To restrict the output of estimated NDV, we limit $\sum_{j=1}^{2k}\Lambda_j=1$ and estimate NDV by exploiting the base estimators:

\begin{align}
    \log \hat{D}=\sum_{j=1}^k(\Lambda_j\cdot\log\hat{\mathcal{D}}|_{\mathcal{I}^\mathrm{over}_j} + \Lambda_{k+j}\cdot\log\hat{\mathcal{D}}|_{\mathcal{I}^\mathrm{under}_j}),
    \label{eq:logd}
\end{align}
where $\hat{\mathcal{D}}|_{\mathcal{I}^\mathrm{over}_j}$ is the estimated NDV of the $j$-th chosen overestimated base estimator, and the output estimated NDV $\hat{D}=e^{\log \hat{D}}$. 
The logarithm is applied to limit the estimation of base estimators so that they do not exceed the range of a 32-bit floating-point number, thus preventing potential impacts on model training.
The learned model solely generates the weight vector, different from previous works that directly estimate NDV.

\subsubsection{Fusion Component Training} Denote $D_i$ as the ground truth NDV of the $i$-th training sample, and the learned model can be trained using:

\begin{align}
    \mathcal{L}_{\mathrm{est}}=\frac{1}{\mathcal{N}}\sum_{i=1}^\mathcal{N}(\log \hat{D}_i-\log D_i)^2+\lambda||W||_2,
    \label{eq:loss-est}
\end{align}
where we apply $L_2$ regularization on model parameters $W$ for better generalization. The regularization parameter $\lambda$ is tuned based on the validation loss.

\section{Experiments}\label{sec:exp}

\subsection{Experimental Setup}\label{sec:exp-settings}
\noindent\textbf{Dataset Selection.} 
A fundamental test case for evaluating estimators is applying them to an individual data column, where the underlying data distribution is manifested. Therefore, the diversity of evaluation scenarios hinges on the variety of individual columns with various data distributions, 
rather than the quantity of data tuples in the data column.
Most traditional statistical estimators~\cite{goodman1949estimation,gee_charikar2000towards,error_bound,chao_in_db_ozsoyoglu1991estimating,chao1984nonparametric,shlosser1981estimation,chaolee,hybskew_haas1995sampling,sichel1986parameter,sichel1986word,sichel1992anatomy,mmo_bunge1993estimating,bootstrap_smith1984nonparametric,horvitz_sarndal1992model,hybskew_haas1995sampling} are tested on a few synthetic columns that satisfy their heuristics and assumptions. 
Recent studies of learned estimators~\cite{li2022sampling,ls_wu2022learning} primarily evaluate the performance on manually crafted standard data distribution (e.g. Zipfian and Poisson) and a limited number of columns of some open-source datasets (SSB~\cite{ssb_o2009star}, Campaign~\cite{Campaign}, NCVR~\cite{NCVR}, et. al.). 
In conclusion, the previous works used limited data distributions and data columns to evaluate the performance of NDV estimators, which may lead to insufficient evaluations.

In recent years, the research community has proposed open-source large-scale tabular datasets~\cite{hulsebos2023gittables,eggert2023tablib}. 
TabLib~\cite{eggert2023tablib}, which collects 627M individual tables totaling 69 TiB, is the largest one all over the world. 
In this paper, we select TabLib sample version~\cite{tablib-v1-sample}, which contains 0.1\% of the full version (69 GB), as our dataset for evaluation. 
TabLib exhaustively contains tabular data from the real world (GitHub~\cite{github} and Common Crawl~\cite{commoncrawl}) with diverse domains,  which can better reflect the data distribution in practice.

\noindent\textbf{Data Preprocess.} TabLib sample version contains 77 parquet files, we remove three of them (2d7d54b8, 8e1450ee, and dc0e820c) for the memory issue. Then we divide the remaining 74 files into train, test, and validation sets to avoid potential data leaks. 
Each parquet file contains thousands of tables, where for each column we independently sample 1\% of data tuples uniformly to construct evaluation cases. 
{Previous works solely focus on large tables with millions of rows~\cite{ls_wu2022learning,li2024learning}. We expand the evaluation cases by assessing the methods on table sizes ranging from tens of thousands to millions of rows.}
The statistics of preprocessed data are shown in Table \ref{tab:data-statistics}, where ``\# Columns'' represents the number of Train/Validation/Test cases used for evaluation.

\begin{table}[]
    \centering
    \caption{Statistics of preprocessed TabLib sample data.}
    \begin{tabular}{ccccccc}
\toprule
        &  Train & Validation & Test     \\
\midrule
         \# Columns  & 89,283 & 30,418 & 25,159  \\

\bottomrule
    \end{tabular}
    \label{tab:data-statistics}
\end{table}

\noindent\textbf{Evaluation Criteria.} To comprehensively evaluate the performance of NDV estimators, we use mean q-error (as defined in Equation~(\ref{eq:q-error})) and the distribution (50\%, 75\%, 90\%, 95\%, and 99\% quantiles) of q-error on a large data volume.

\noindent\textbf{Implementation Details.} We implement our model in PyTorch, and the implementation details are shown as follows: 
the optimizer is Adam~\cite{adam_kingma2014adam} with an initial learning rate of 0.001, $\alpha$ is 1, $\beta$ is 0.5, the number of input features $H$ is 100, the number of selected leading estimators $k$ is 2. The training epoch is 100, we save the model by 99\% quantile of q-error on the validation set and report the performance on the test set. All the experiments in this paper are conducted on an NVIDIA A100 GPU.

\noindent\textbf{Baseline Models.} 
In our evaluation, we include statistical estimators, hybrid estimators that integrate statistical ones, and learned estimators as our baselines.

\noindent\textit{Statistical estimators (base estimators).}
There are many statistical estimators, we select representative ones as our baselines as well as our base estimators.

\begin{itemize}[leftmargin=10pt]
\item Goodman~\cite{goodman1949estimation} is the seminal work in NDV estimation, and we use the expression in Equation (\ref{eq:goodman}). 
\item GEE~\cite{gee_charikar2000towards} provides a theoretical lower bound of ratio error and it uses geometric mean as scale factor: $D_{\mathrm{GEE}}=\sqrt{N/n}f_1+\sum_{j=2}^nf_j$.
\item Error Bound (EB)~\cite{error_bound} is proposed to estimate NDV in sampling-based histogram construction, and $D_{\mathrm{EB}}=\sqrt{N/n}f_1^++\sum_{j=2}^nf_j,f_1^+=\max(1,f_1)$.
\item Chao~\cite{chao1984nonparametric,chao_in_db_ozsoyoglu1991estimating} assumes the size of $C$ is infinity. We use the expression in Equation (\ref{eq:chao}). If $f_2$ is zero, we will return $d$.
\item Shlosser~\cite{shlosser1981estimation} is based on the assumption that the frequency profile of sample data is approximately the frequency profile of the original column. ${D}_{\text{Shlosser}} = d + \frac{f_1 \sum_{i=1}^n (1 - r)^i f_i}{\sum_{i=1}^n ir(1 - r)^{i-1} f_i}$.
\item ChaoLee~\cite{chaolee} adds another estimator in Chao to treat data skew, we refer to~\cite{ndvlib} to implement it.
\item Jackknife~\cite{burnham1978estimation,burnham1979robust} assumes $d_n$ be the NDV of the sample and numbers the tuples from 1 to n in the sample data. Denote $d_{n-1}(k),1\leq k \leq n$, $d_{n-1}(k)=d_n-1$ if the attribute value for tuple $k$ is unique; otherwise $d_{n-1}(k)=d_n-1$. The first-order Jackknife estimator is: $D_{\mathrm{Jackknife}}=d_n-(n-1)(d_{n-1}-d_n)$.
\item Sichel~\cite{sichel1986parameter,sichel1986word,sichel1992anatomy} estimator needs to solve non-linear equations, as shown in Equation~(\ref{eq:sichel}).
\item Bootstrap~\cite{bootstrap_smith1984nonparametric}: $D_{\mathrm{Boot}}=d+\sum_{j:n_j>0}(1-n_j/n)^n$. It may perform worse when $D$ is large and $n$ is small because $D_{\mathrm{Boot}}\leq 2d$.
\item Horvitz-Thompson (HT)~\cite{horvitz_sarndal1992model} has a sophisticated expression, it defines $h_n(x)=\frac{\Gamma(N-x+1)\Gamma(N-n+1)}{\Gamma(N-n-x+1)\Gamma(N+1)}$, where $\Gamma$ is the gamma function, and $D_{HT}=\sum_{j:n_j>0}\frac{1}{1-h_n(\hat{N}_j)},\hat{N}_j=N(n_j/n)$. 
\item Method of Movement (MoM)~\cite{mmo_bunge1993estimating} has three versions. MoM v1 assumes the frequencies are equal ($N_1=N_2=\ldots=N_D$) and an infinite population, it needs to solve the equation: $d=D_{\mathrm{MoM1}}(1-e^{-n/D_{\mathrm{MoM1}}})$. MoM v2 assumes the population size is finite, and the estimator is $d=D_{\mathrm{MoM2}}(1-h_n(N/D_{\mathrm{MoM2}}))$. MoM v3 assumes the frequencies are unequal and it has a sophisticated expression, we refer to~\cite{ndvlib} to implement it.
\item Smoothed Jackknife (SJ)~\cite{hybskew_haas1995sampling}. $D_{\mathrm{SJ}}=d_n-K((d_{n-1}-d_n))$, there is a extremely sophisticated approximation expression for $K$, we omit its expression in the paper and refer to ~\cite{ndvlib} to implement it.
\end{itemize}

\noindent\textit{Hybird estimators.} Existing hybrid estimators are commonly constructed by using SJ~\cite{hybskew_haas1995sampling}, GEE~\cite{gee_charikar2000towards}, Shlosser~\cite{shlosser1981estimation} estimators. Specifically, they use $\chi^2_{n-1}$ test~\cite{chi2test} to pick estimators. Define 
\begin{align}
u=\sum_{j,n_j>0}(\frac{(n_j-\Bar{n})^2}{\Bar{n}}),\Bar{n}=\frac{n}{d}.
\end{align}
\begin{itemize}[leftmargin=10pt]
\item HYBSkew~\cite{hybskew_haas1995sampling} uses SJ estimator if $u\leq \chi^2_{n-1,0.975}$, otherwise it takes Shlosser estimator.
\item HYBGEE~\cite{gee_charikar2000towards} uses SJ estimator if $u\leq \chi^2_{n-1,0.975}$, otherwise it takes GEE estimator.
\end{itemize}

\noindent\textit{Learned estimators.}
In addition, we compare our method with the open-sourced SOTA learned estimator LS~\cite{ls_wu2022learning}. We consider three variants of LS for a fair comparison:

\begin{itemize}[leftmargin=10pt]
\item LS$_{\mathrm{general}}$: since LS claimed can directly apply to any data distribution~\cite{ls_wu2022learning}, we use its open-source checkpoint as a baseline.
\item LS$_{\mathrm{scratch}}$: we train LS from scratch on the same training set as \textsc{AdaNDV}.
\item LS$_{\mathrm{FT}}$: we fine-tune (FT) LS$_{\mathrm{general}}$ as described in~\cite{ls_wu2022learning} on the same training set as \textsc{AdaNDV}.
\end{itemize}

Besides, we construct two learned estimator baselines as follows:

\begin{itemize}[leftmargin=10pt]
\item Select-Optimal (SO): it selects one optimal base estimator using an MLP with the same architecture as the estimator selection model in \textsc{AdaNDV}, designed to accomplish the objective as Hypo-optimal.
\item Learnable Ensemble (LE): it integrates the results of all fourteen base estimators by a learnable weighted sum, where the number of parameters is equal to the number of base estimators.
\end{itemize}

\begin{table}[]
    \centering
    \caption{Mean and quantiles of q-error of baselines and our method, where $\infty$ indicates that the number exceeds the representation limits of a 32-bit floating-point type. Each best-performing metric is emphasized in boldface.}
    \begin{tabular}{ccccccc}
\toprule
        Estimator& Mean & 50\% & 75\%& 90\% & 95\% & 99\% \\
\midrule
        Goodman& $\infty$ &4.14 & 37.91& 100.78&1.11e11 & $\infty$  \\
        GEE  &3.71  &1.97  &3.86  &9.91  &10.10  &11.70\\
        EB & 3.98 & 2.62 & 6.00 & 9.98 & 10.12 & 11.00 \\
        Chao & 21.98 & 1.85 & 6.50 & 99.99 & 100.04 & 100.19 \\
        Shlosser & 4.25 & 1.80 & 4.53 & 10.41 & 15.14 & 25.14\\
        ChaoLee &   23.28 & 4.35 & 29.39 & 96.00 & 99.36 & 100.51 \\
        Jackknife & 12.60 & 2.80 & 16.24 & 49.04 & 50.38 & 51.71  \\
        Sichel & 152.93 & 2.49 & 99.70 & 365.10 & 1.03e3 & 2.20e3 \\
        MoM v1 & 2.51e4 & 2.00 & 6.60 & 22.70 & 66.51 & 1.11e6 \\
        MoM v2 & 8.60 & 3.14 & 9.96 & 18.84 & 30.61 & 86.09 \\
        MoM v3 & 4.29e3 & 5.95 & 61.92 & 818.18 & 3.30e3 & 4.30e4 \\
        Bootstrap &  70.01 & 10.99 & 47.91 &95.42 &224.09& 1.25e3 \\
        HT & 2.83e3& 41.56 & 354.42 & 5.07e3 & 1.03e4 & 4.41e5 \\
        SJ & 231.67 & 2.17 & 8.12 & 34.03 & 90.25 & 1.81e3\\
\midrule
        HYBSkew & 4.25 & 1.80 & 4.53 & 10.41 & 15.14 & 25.14\\
        HYBGEE & 3.71 & 1.97 & 3.86 & 9.91 & 10.10 & 11.70\\
\midrule
    LS$_{\mathrm{general}}$ & 2.24 & 1.72 & 2.28 & 3.20 & 4.11 & 10.46 \\

    LS$_{\mathrm{scratch}}$ & 1.91 & 1.36 & 1.80 & 2.57 & 3.53 & 10.80\\
    
    LS$_{\mathrm{FT}}$ & 1.96 & 1.39 & 1.86 & 2.64 & 3.63 & 11.44 \\

    {SO}  & $\infty$ & 1.29 & 2.08 & 3.29& 4.42 & 13.03 \\
    
    {LE} & 2.30 & 1.71 & 2.22 & 3.44 & 4.63 & 11.88\\

\midrule
    \textsc{AdaNDV} & \textbf{1.62} & \textbf{1.22}& \textbf{1.60} & \textbf{2.34} & \textbf{3.24} & \textbf{6.79} \\
    
\bottomrule
    \end{tabular}
    \label{tab:overall}
\end{table}

\subsection{Statistical Estimators Analysis}
\label{sec:perf-traditional}
In this section, we present a detailed analysis of the performance of statistical and hybrid estimators in large-scale real-world scenarios.

\noindent\textbf{Overall Performance.} The performance is shown in Table \ref{tab:overall}. According to the results, we can draw the following conclusions:

\begin{itemize}[leftmargin=10pt]
\item There is no single traditional estimator that always outperforms the other ones across all the evaluated metrics. 
\item The majority of traditional estimators exhibit subpar performance across most metrics. In the most adverse scenarios, the q-error of the Goodman estimator even surpasses the upper limit that a 32-bit floating-point value can represent due to factorial operations in Equation~(\ref{eq:goodman}).
\item Using a comprehensive assessment of the performance of NDV estimators is desired. 
For example, although the Shlosser estimator has a mean q-error that is higher but close to that of EB, its 50\% quantile of q-error is considerably lower, while its 99\% quantile of q-error is significantly higher. A single metric can not reflect the performance of an estimator.
\item HYBSkew does not outperform Shlosser, and HYBGEE does not surpass GEE. On the one hand, they rarely select SJ, indicating hybrid estimators may mitigate poor results to some extent. On the other hand, the performance of traditional hybrid estimators is limited by the selected single estimator.
\end{itemize}

\noindent\textbf{Intrinsic Weaknesses.} The intrinsic weaknesses of statistical estimators are evident and have been discussed in previous works~\cite{ls_wu2022learning,li2024learning}: the assumptions and conditions of traditional estimators are infrequently satisfied in practice.

We expand the number of evaluation cases by several orders of magnitude to provide a more comprehensive assessment that has not been investigated before, and we can conclude that the weaknesses exposed by prior studies remain valid.

\noindent\textbf{Undiscovered Strengths.} 
However, we argue that the strengths of statistical estimators are significantly eclipsed by their weaknesses. Although real-world data distributions often fail to meet their presupposed conditions, it is possible to identify estimators that deliver the estimated result with an acceptable level of q-error. 
Our argument is supported by the Hypo-optimal estimator shown in Table \ref{tab:ideal-traditional}. 
Meanwhile, \textit{it suggests that accurately selecting appropriate estimators could significantly reduce q-error}.

\begin{figure}
    \includegraphics[width=\linewidth]{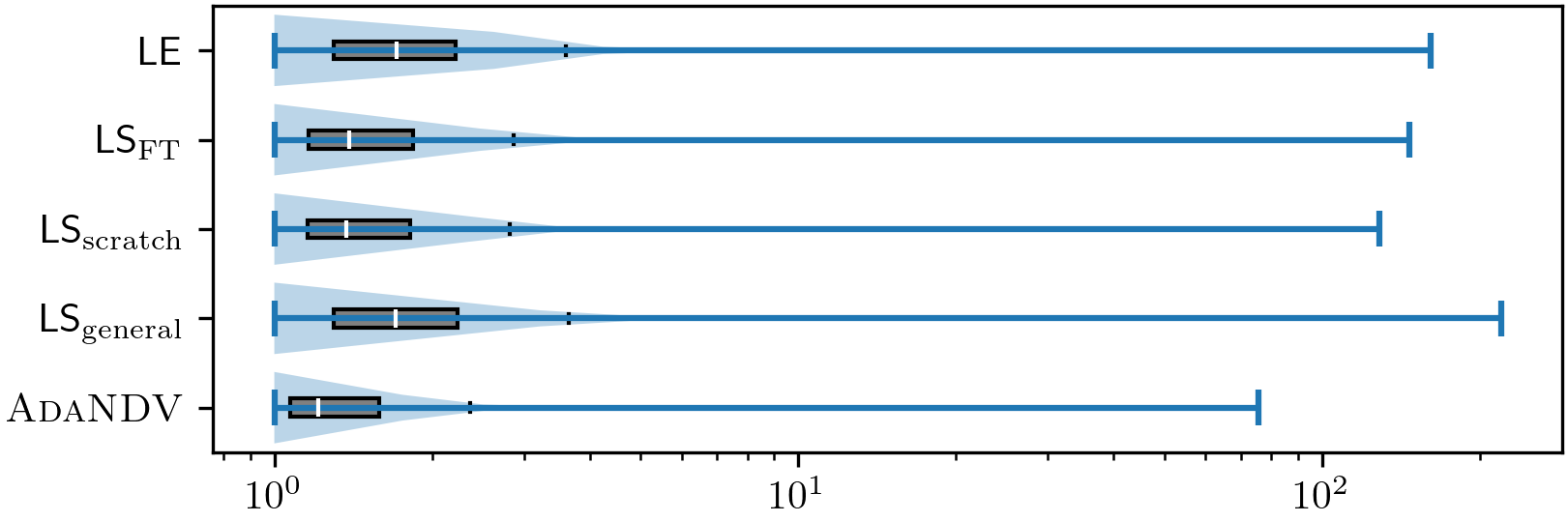}
    \centering
    \caption{Error distribution of learned estimators on the test set. The violin plot is in blue. The boxplot is in black, the gray box contains 50\% of data points, and the white line in the gray box represents the median. We exclude the SO estimator due to its extremely large mean error.}
    \label{fig:err-distribution}
\end{figure}

\subsection{Results of Learned Estimators}
\label{sec:perf-adandv}

\noindent\textbf{Effectiveness.} The overall performance of learned estimators is shown in Table \ref{tab:overall}, and we observe the following findings:

\begin{itemize}[leftmargin=10pt]
\item \textit{Performance of the proposed learned estimator}.
\textsc{AdaNDV} significantly outperforms all estimators across all the metrics. These results substantially demonstrate the superiority of \textsc{AdaNDV}.

\item \textit{Performance of SOTA learned estimators}. 
LS$_{\mathrm{general}}$ consistently outperforms the statistical estimators across all the metrics, which exhibits the advantages of learned estimators that can adapt to data shifting over statistical estimators.
In addition, LS$_{\mathrm{scratch}}$ and LS$_{\mathrm{FT}}$ have better performance compared to LS$_{\mathrm{general}}$ in general.
However, by looking at the different quantiles of q-error, training or fine-tuning may decline the robustness, where they perform worse than LS$_{\mathrm{general}}$ for 99\% quantile of q-error.

\item \textit{Performance of constructed learned estimators}. The performance of the SO and LE estimators reveals the necessity of investigating the historically overlooked issues in NDV estimation. Firstly, selecting one optimal estimator by a learned model can outperform individual estimators in most scenarios, but it is challenging to achieve high accuracy, which can lead to worse cases. Besides, the LE estimator outperforms individual statistical ones, showcasing the potential of estimator fusion.

\item \textit{Error distribution discussion}. 
In Table \ref{tab:overall}, we also observe that \textsc{AdaNDV} exhibits a more robust error distribution: about 90\% of the test cases show a q-error below 2.30, and approximately 99\% of test cases exhibit the q-error do not exceed 7. 
Moreover, we depict the boxplot~\cite{boxplot} and violin plot~\cite{violinplot} of q-error distributions of learned estimators in Figure~\ref{fig:err-distribution}, where we omit the SO estimator for its overflow issue. 
From the figure, we observe that the q-error distribution for the AdaNDV shows a higher concentration around values closer to the minimum error, indicating its advantages. Besides, the maximal q-error of \textsc{AdaNDV} is much smaller than other learned estimators, representing better performance in the worst cases.

\end{itemize}

\noindent\textbf{Efficiency.} 
The efficiency of an estimator determines its practicality. 
Thus we also evaluate the efficiency of $\textsc{AdaNDV}$ and the learned estimators in terms of both time and space consumption.
The computing overhead of $\textsc{AdaNDV}$ involves the neural networks and the selected base estimators. 
Since some base estimators involve solving non-linear equations, it is not easy to provide a rigorous time complexity analysis. 
Therefore, for simplicity, we record the end-to-end latency of the training and testing stages for them.

Specifically, in the training stage, we record the execution time of the base estimators as well as the time reaching convergence for the learned estimators. 
In the inference stage, we capture the time required for processing all test cases. 
In addition, we compare the learnable parameters between the learned estimators to show their space efficiency. We exclude LS$_{\mathrm{general}}$ in this comparison since it is pre-trained and has the same inference efficiency and parameters as its variants.
The time and space consumptions are shown in Table \ref{tab:efficiency}, and we derive the following conclusions.

\begin{table}[]
    \centering
    \caption{Efficiency comparison of the methods.}
    \begin{tabular}{ccccccc}
\toprule
         & LS$_{\mathrm{scratch}}$ & LS$_{\mathrm{FT}}$ & \textsc{AdaNDV} & SO & LE \\
\midrule
        Train (s) & 929 & 2,963 & {712} & 1,261 & 1,337\\
        Inference (s)   & 25.42 & 25.42 & 51.44 &  32.79  & 84.85 \\
        \# Params& 62,435 & 62,435 & 55,328 & 22,294  & 14  \\
\bottomrule
    \end{tabular}
    \label{tab:efficiency}
\end{table}

\begin{itemize}[leftmargin=10pt]
\item \textit{Training efficiency of learned estimators.} 
We observe that LS$_{\mathrm{FT}}$ is the most time-consuming method in that it requires about three times of epochs than LS$_{\mathrm{scratch}}$ to converge. 
One possible reason may be that adapting the general model to the training domain requires more time than learning from scratch since the fine-tuning process may need to maintain the features learned in the pre-training stage.  
{\textsc{AdaNDV} requires minimal time for convergence.}
The SO and LE estimators have longer training time than \textsc{AdaNDV} and LS$_{\mathrm{scratch}}$, indicating their objectives are more difficult to converge.

\item \textit{Inference efficiency.} 
The inference overhead of baseline learned estimators solely involves a neural network model inference and the total time on the test set ($25,159$ samples) is 25.42s, the average inference time of a test case is about \textbf{1ms}. 
In {contrast}, \textsc{AdaNDV} consists of three neural models and they need 27.46s to finish the inference on the test set. 
Besides, \textsc{AdaNDV} additionally requires the base estimator computation which needs 23.97s. 
In total, \textsc{AdaNDV} spends 51.44s to finish the inference on all the test samples with an average estimation time of about \textbf{2ms}.
The elaborated components bring performance improvement and efficiency decline in \textsc{AdaNDV}.
For the constructed estimators, the SO estimator requires a selected base estimator employment for each test case, resulting in a longer inference time compared to LS models. LE needs the estimation results of all base estimators, leading to the longest inference time among the learned estimators.

\item \textit{Space efficiency comparison.} 
We observe that the variants of LS possess approximately 12.85\% more parameters compared to \textsc{AdaNDV}.
Its advantage can likely be attributed to the elaborated integration of base estimators. 
This strategic utilization enhances the performance of learned estimator \textsc{AdaNDV} with less representation ability (number of learnable parameters). 
For the constructed estimators, the SO estimator has fewer parameters than \textsc{AdaNDV} and LS models, since its architecture is the same as a component in \textsc{AdaNDV}. LE solely has 14 learnable parameters, making it the most lightweight among the learned estimators, which may indicate the potential for lightweight optimization in \textsc{AdaNDV}.

\end{itemize}

\begin{table}[t]
    \centering
    \caption{Ablation study, where ``{w/o}'' indicates removing a component from \textsc{AdaNDV}.}
    \begin{tabular}{ccccccc}
\toprule
        Estimator& Mean & 50\% & 75\%& 90\% & 95\% & 99\% \\
\midrule
         \textsc{AdaNDV} & {\textbf{1.62}} & {\textbf{1.22}}& {\textbf{1.60}} & {\textbf{2.34}} & {\textbf{3.24}} & {\textbf{6.79}} \\
        w/o select  & {2.30} & {1.71} & {2.22} & {3.44} & {4.63} & {11.88
} \\ 
        w/o fusion  &  {2.42} & {1.35} & {2.01} & {3.31} & {4.75} & {19.30} \\
\midrule
        {w/o} log & 2.72e7 &  {41.54}& {671.06} & {2.67e3} & {5.87e4}  & 1.00e10 \\
         {w/o} base & {1.67}  & {1.27}  & {1.66} & {2.39}  & {3.34}  & {7.42}  \\
         {w/o} over & {2.20} & {1.37} & {1.92} & {3.10} & {4.27} & {11.97} \\
         {w/o} under & {2.49} & {1.82} & {2.88} & {5.50} & 6.00 & 11.00  \\
         {w/o} comp & {2.05} & 1.26 & 1.76 & {2.68} & {3.81} & {11.68} \\
\bottomrule
    \end{tabular}
    \label{tab:ablation}
\end{table}

\subsection{Ablation Study}
\label{sec:exp-ablation}
The outstanding performance of \textsc{AdaNDV} mainly stems from the effective collaboration between its leading estimator selection and the fusion-based estimation.
To investigate the contributions of each module, we develop several variants of \textsc{AdaNDV}. The symbol ``{w/o}'' indicates removing the component or strategy from our method.

\noindent\textbf{Primary Components.} 
We individually eliminate each of the two principal constituents within \textsc{AdaNDV}.
Firstly, we drop the leading estimator selection component to deliberately make the model use all base estimators to estimate NDV, the variant is named ``\textbf{{w/o} select}''. 
We then remove the estimator fusion component to select a single estimator by the ranking paradigm from the base estimators to present NDV, the variant is named ``\textbf{{w/o} fusion}''. 
The performances of the two variants are shown in Table \ref{tab:ablation}. 
We can observe that the performance of \textsc{AdaNDV} significantly declines across all the metrics if we remove one of the two components. Based on the observation, we can derive the following conclusions. 
\begin{itemize}[leftmargin=10pt]
\item Directly estimating NDV by fusing all base estimators may make the task difficult because it has to leverage poor-performing estimators. Besides, the performance of ``{w/o} select'' surpasses that of LE, highlighting the advantages of adaptively adjusting the weights of fused estimators for different test cases.
\item The performance of ``{w/o} fusion'' shows that selecting a single estimator with high accuracy is challenging because of the representation ability (sparse features and the MLP architecture) of the model. On the contrary, it performs better than the SO estimator in most metrics and can alleviate the $\infty$ of mean q-error. It demonstrates the effectiveness of the ranking paradigm (``{w/o} fusion'') over the classification paradigm (SO), and further investigation will be conducted in Section~\ref{sec:exp-paradigm}.
\item Each component has a significant contribution to \textsc{AdaNDV}, and the combination of two components leads to the superiority of \textsc{AdaNDV}. 
\end{itemize}

\noindent\textbf{Detailed Strategies.} 
We further study the other strategies utilized in \textsc{AdaNDV} and the performance of dropping each strategy is shown in Table \ref{tab:ablation}. 
Firstly, we remove the log scale training technique by transforming Equation (\ref{eq:logd}) to $\hat{D}=\prod_{j=1}^k(\hat{\mathcal{D}}|_{\mathcal{I}^\mathrm{over}_j}^{\Lambda_j}\cdot \hat{\mathcal{D}}|_{\mathcal{I}^\mathrm{under}_j}^{\Lambda_{k+j}})$, the variant is denoted as ``\textbf{{w/o} log}''. 
The performance of ``{{w/o} log}'' significantly declined, which indicates the effectiveness of logarithm operation.

In the estimator fusion component, we add the estimated NDV of the selected leading estimators as features. 
We remove the features of estimated results, and the variant is named ``\textbf{{w/o} base}''. 
Its performance slightly declined on all metrics but performs better than all baselines, which indicates that taking the estimated results as input features may make the model aware of the input value domain and then improve the weights allocation to get more accurate estimations.

\begin{figure*}[t]
     \centering
    \begin{subfigure}{0.16\textwidth}
         \centering
         \includegraphics[width=\textwidth]{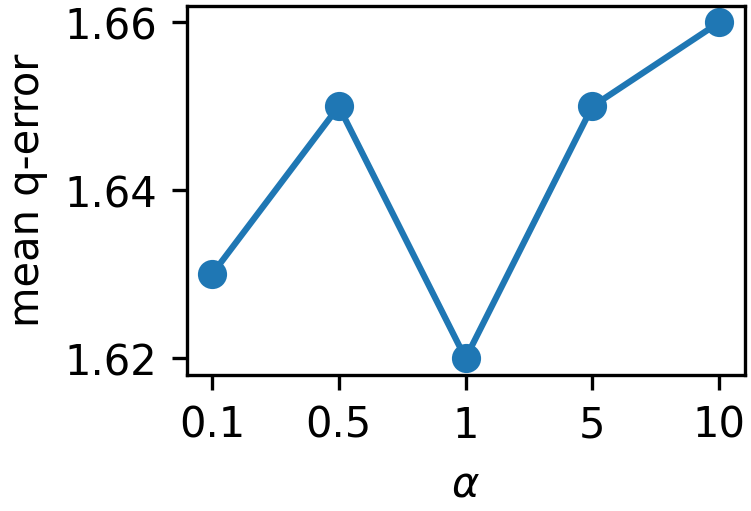}
     \end{subfigure}
     \hfill
     \begin{subfigure}{0.16\textwidth}
         \centering
         \includegraphics[width=\textwidth]{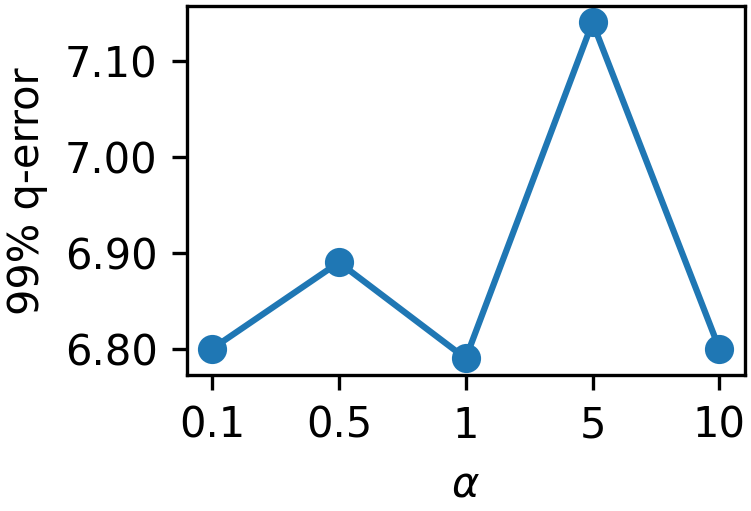}
     \end{subfigure}
     \hfill
     \begin{subfigure}{0.16\textwidth}
         \centering
         \includegraphics[width=\textwidth]{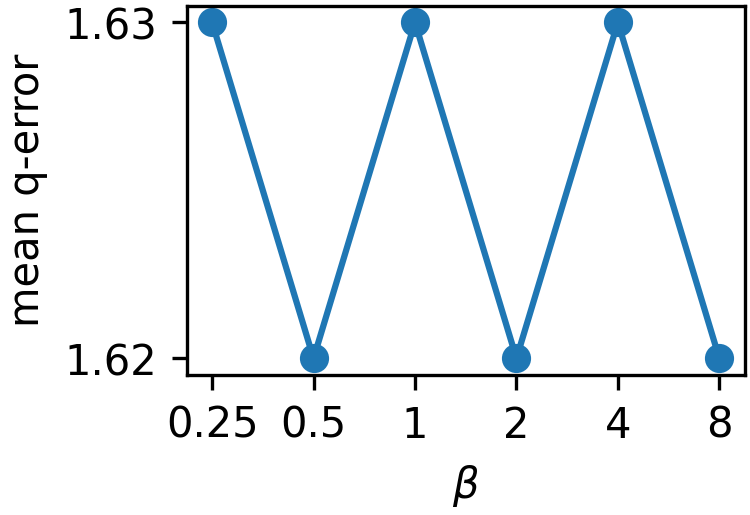}
     \end{subfigure}
     \hfill
     \begin{subfigure}{0.16\textwidth}
         \centering
         \includegraphics[width=\textwidth]{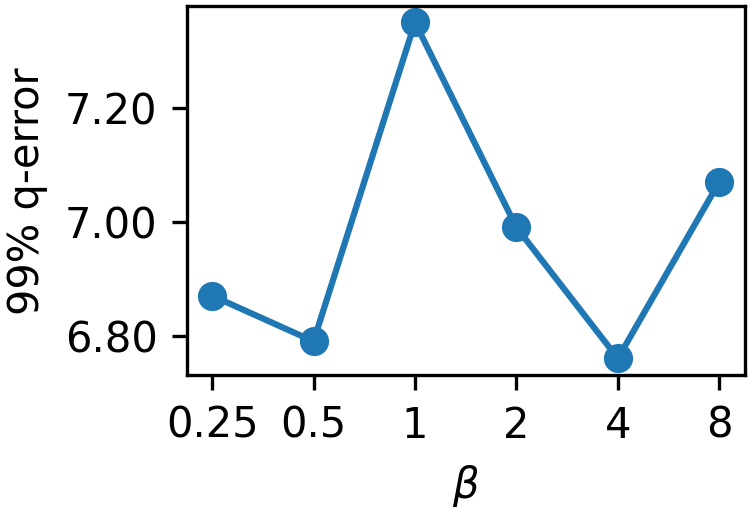}
     \end{subfigure}
     \hfill
     \begin{subfigure}{0.16\textwidth}
         \centering
         \includegraphics[width=\textwidth]{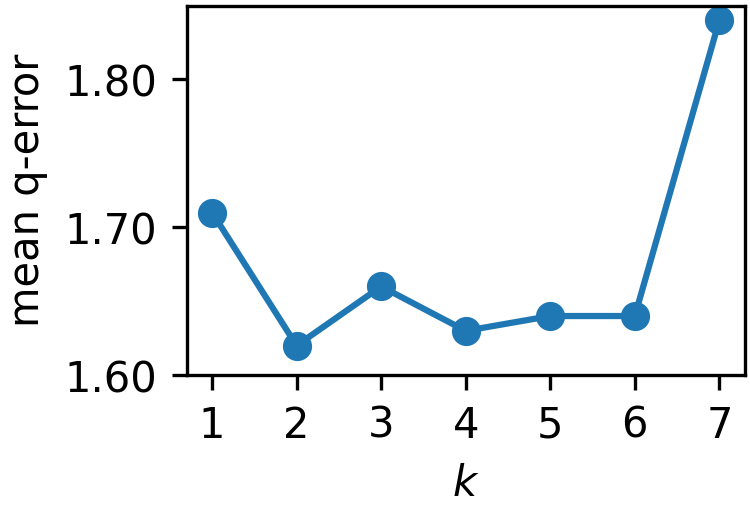}
     \end{subfigure}
     \hfill
     \begin{subfigure}{0.16\textwidth}
         \centering
         \includegraphics[width=\textwidth]{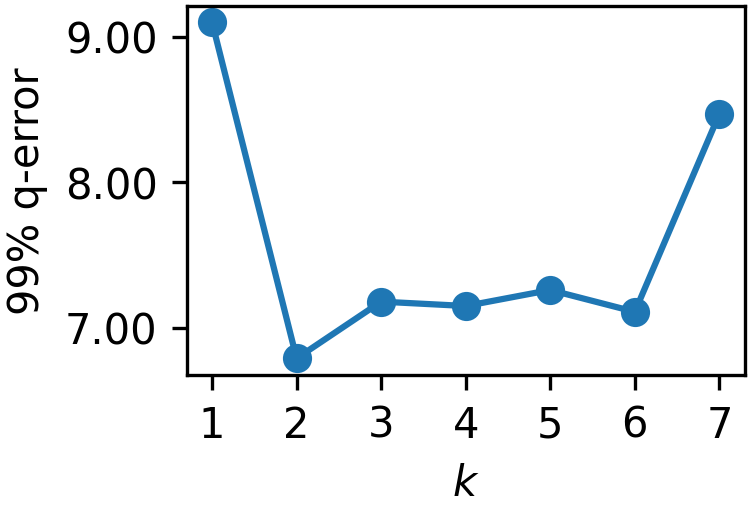}
     \end{subfigure}
     \hfill
    \caption{Performance on mean and 99\% percentile of q-error of \textsc{AdaNDV} with different hyperparameters.}
    \label{fig:hyper-all}
\end{figure*}

Finally, we study whether discerning overestimation and underestimation estimators work. 
We remove the two kinds of leading estimators and name the two variants as ``\textbf{{w/o} over}'' and ``\textbf{{w/o} under}'', respectively. 
We can observe the two variants significantly decrease on all metrics and they cannot beat baseline learned estimators. 
The performance illustrates that individually utilizing the two kinds of leading estimators can not obtain satisfactory results but the combination of them can bring a significant approximation to the ground truth NDV. In addition, we construct a new variant ``\textbf{{w/o} comp}'' that directly selects {2$k$} estimators without deliberately utilizing the overestimation-underestimation complementary error correction strategy. It outperforms all variants in most metrics but is consistently worse than \textsc{AdaNDV}. On the one hand, it illustrates the benefit of our proposed complementary error correction strategy. On the other hand, although not intentionally distinguished, it is rare for the selected estimators to be exclusively overestimated or underestimated. This further demonstrates the effectiveness of our strategy combined with the experimental conclusions of ``{w/o} over'' and ``{w/o} under''.

\subsection{Impact of Hyperparameters}
\label{sec:exp-hyperparameter}
There are three hyperparameters in \textsc{AdaNDV}: $\alpha, \beta, k$, and we investigate the sensitivity of the model. We demonstrate the performance of mean and 99\% percentile of q-error in Figure~\ref{fig:hyper-all} to show the impact of hyperparameters.

\noindent\textbf{Effect of $\alpha$ in the Ranking Model.} We can observe different $\alpha$ values lead to some fluctuations in the performance, but \textsc{AdaNDV} is not sensitive to different values of $\alpha$. We set $\alpha$ to 1 because it achieves the best performance in most metrics.
Extensively discussing the effect of this knob is out of the scope of this paper, refer to~\cite{bruch2019revisiting,wang2018lambdaloss} for more details.

\noindent\textbf{Effect of Multi-objective Optimization Knob $\beta$.} As shown in Figure~\ref{fig:hyper-all}, we observe that the performance in mean q-error remains unchanged across different values of the knob $\beta$. However, there is a notable difference in 99\% q-error. For each experimented value of $\beta$, \textsc{AdaNDV} exhibits relatively stable performance. The knob $\beta$ balances the estimator ranking and NDV prediction tasks, we set $\beta$ as 0.5 in this paper because it achieves better results in most metrics than in other settings.

\noindent\textbf{The Number of Leading Estimators $k$.} The number of selected leading estimators directly affects the input of the estimation fusion component. With the variation of $k$, the performance of the model does not improve beyond that achieved with $k=2$ on the mean and the 99\% percentile of q-error. This finding illustrates that selecting more estimators does not consistently benefit the fusion component, as demonstrated in the analysis of the LE estimator and the variant of \textsc{AdaNDV} ``\textbf{{w/o} select}''. Additionally, increasing $k$ involves incorporating more base estimators during the inference stage, which introduces additional computational overhead, as shown in Table~\ref{tab:efficiency}. Considering both effectiveness and efficiency, we set $k$ as 2.

\subsection{Flexibility Discussion}
\label{sec:exp-paradigm}

\begin{figure}[t]
    \centering
    \begin{subfigure}{0.23\textwidth}
         \centering
         \includegraphics[width=\textwidth]{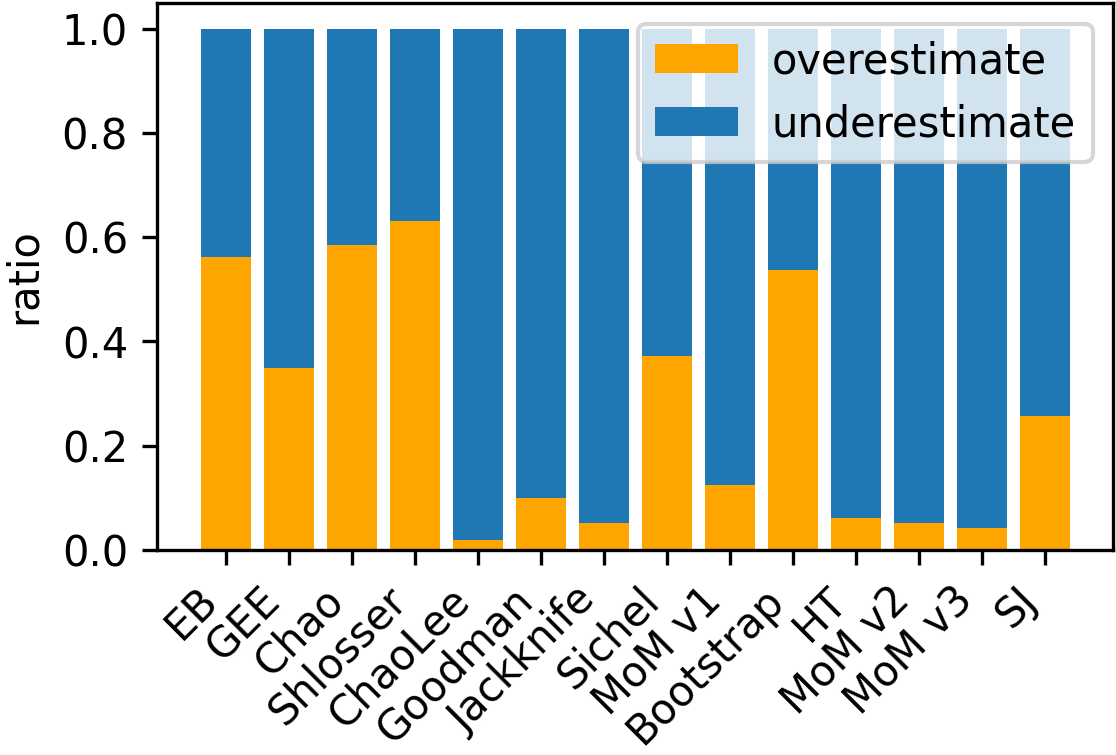}
         \caption{{Performance of base estimators.}}\label{fig:baseratio}
     \end{subfigure}
     \begin{subfigure}{0.23\textwidth}
         \centering
         \includegraphics[width=\textwidth]{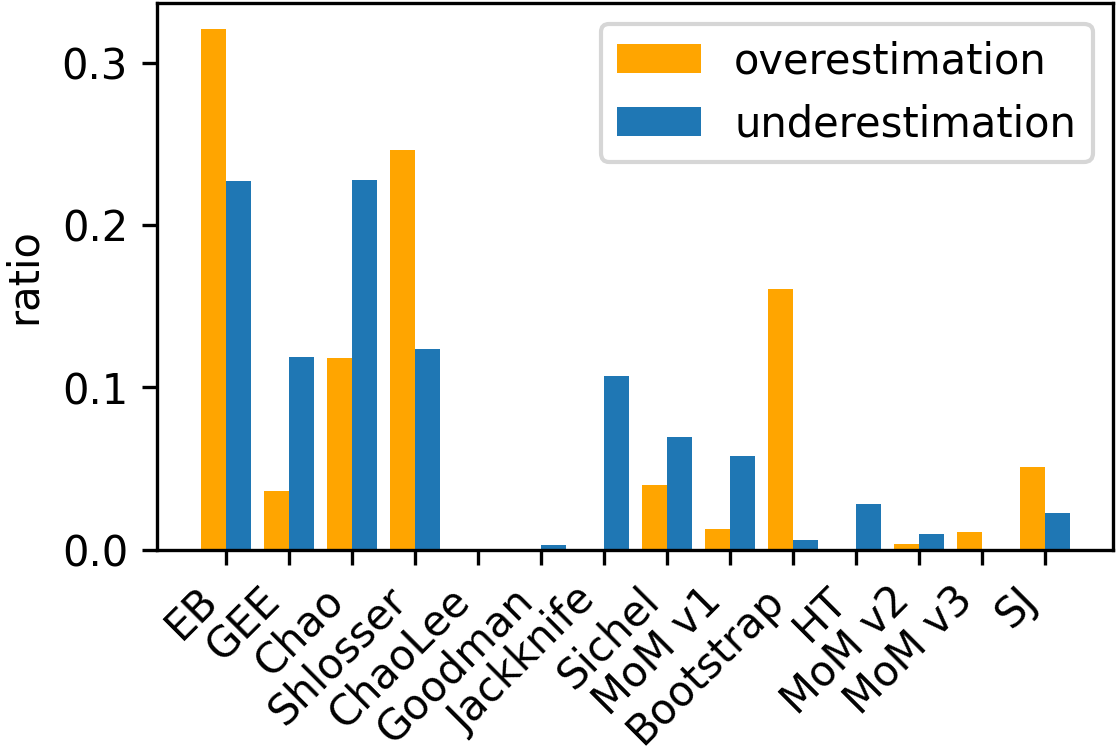}
         \caption{Proportion of selected estimators.}\label{fig:chosen}
     \end{subfigure}
     \caption{{Illustration of overestimation and underestimation properties.}}\label{fig:overunderproperty}
\end{figure}

\noindent\textbf{Selected Estimator Discussion.} 
{Since we leverage the property of overestimation and underestimation, we demonstrate the overestimation and underestimation ratios of base estimators in Figure~\ref{fig:baseratio}.} We count the estimators selected by the overestimated and underestimated selection models of \textsc{AdaNDV}, and the proportion of each estimator is shown in Figure~\ref{fig:chosen}. We can observe the following findings based on the results.

\begin{itemize}[leftmargin=10pt]
\item {Base estimators can hardly derive the ground truth, as they always either overestimate or underestimate. Besides, no estimators can consistently overestimate or underestimate, but certain estimators tend to underestimate on the dataset.}
\item EB is the most frequently selected by both models, ChaoLee has not been selected by either model, and some estimators are exclusively selected by a single model. It suggests that most estimators are beneficial and the overestimation-underestimation complementary perspective is utility. 
\item Figure~\ref{fig:traditional-strength} illustrates that all base estimators can achieve the lowest q-error in certain scenarios compared to others, but some are selected infrequently. This highlights that collecting the set of base estimators may be an open research question.  
\end{itemize}

\noindent\textbf{{Precisions of} Estimator Selection.}
We have constructed the SO estimator as our baseline {by categorizing} the estimators into two classes {for training}: the optimal overestimated or underestimated estimator and otherwise. To further investigate the differences between the ranking and classification paradigms, we develop another variant of \textsc{AdaNDV} by transforming the SO estimator as the estimator selection component in \textsc{AdaNDV}, and the method is named \textsc{AdaNDV}(C). The performance is shown in Table~\ref{tab:flex}, and we can derive the following conclusion. {On the one hand,} {AdaNDV}(C) consistently outperforms all baseline estimators, demonstrating the necessity of treating the selection dilemma problem. {On the other hand,} \textsc{AdaNDV} surpasses \textsc{AdaNDV}(C) on all metrics. The two variants have identical estimator fusion components, indicating that selected estimators directly affect the ultimate estimations.

\begin{table}[t]
    \centering
    \caption{Performance comparison of different variants of \textsc{AdaNDV}.}
    \begin{tabular}{ccccccc}
\toprule
         & Mean & 50\% & 75\%& 90\% & 99\% \\
\midrule
    \textsc{AdaNDV} & {{1.62}} & {{1.22}}& {{1.60}} & {{2.34}} & {{6.79}} \\
    \textsc{AdaNDV}(C) & {1.86} & {1.34} & {1.98}  & {3.04} & {8.42} \\
    \textsc{AdaNDV}{(base}-EB{)} & {1.64} & 1.24 & 1.62 & {2.39} & {7.11}\\
    \textsc{AdaNDV}{(base}+LS$_\mathrm{general}${)} & 1.62 & 1.21 & 1.58 & 2.33 & {7.18} \\
    \textsc{AdaNDV}(Hypo) & 1.18 & {1.10} & 1.24 & {1.43} & {2.20} \\
\bottomrule
    \end{tabular}
    \label{tab:flex}
\end{table}

\begin{figure}[t]
    \centering
    \begin{subfigure}{0.21\textwidth}
         \centering
         \includegraphics[width=\textwidth]{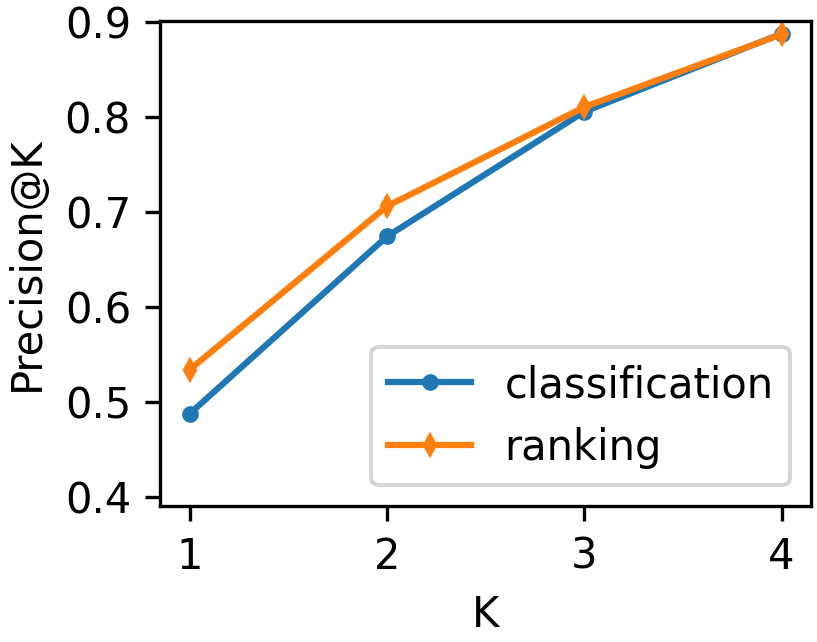}
    \caption{Selection precisions.}
    \label{fig:cmp-paradigm}
     \end{subfigure}
     \begin{subfigure}{0.25\textwidth}
         \centering
         \includegraphics[width=\textwidth]{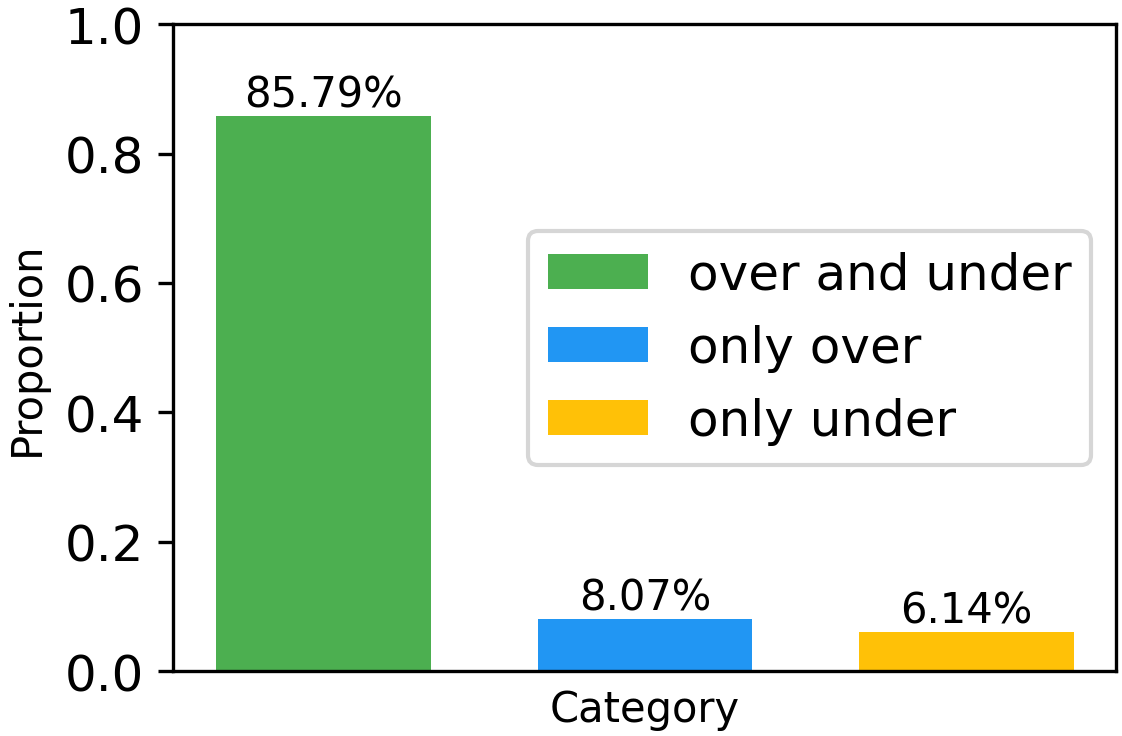}
         \caption{{Type of selected estimators.}}\label{fig:cases}
     \end{subfigure}
     \caption{{Analysis of estimator selection.}}\label{fig:impact}
\end{figure}

We use the metric \textit{Precision@K} (P@K) to evaluate the accuracy {of estimator selection}, representing the partition of the optimal estimator(s) amongst the top-K estimators, and the results are shown in Figure~\ref{fig:cmp-paradigm}. We can observe the model trained in the ranking paradigm exhibits substantially higher P@1 and P@2 than that trained in the classification paradigm.
The sensitive analysis results in Figure~\ref{fig:hyper-all} show that the ultimate performance is closely related to the top estimators and we set $k$ as 2 for \textsc{AdaNDV}, so P@1 and P@2 of leading estimator selection directly affect the performance of our exploitation.

\noindent\textbf{{Selected Estimators for Fusion.}}
{We summarize the fusion scenarios of over/under estimation properties of the 2$k$ ($k=2$) selected estimators in Figure~\ref{fig:cases}. The term ``over and under'' indicates that both overestimated and underestimated results exist among the 2$k$ estimators, meaning their weighted sum can encompass the ground truth. Conversely, ``only over'' and ``only under'' refer to the selected estimators that exhibit only overestimation or underestimation, respectively. In most test cases, \textsc{AdaNDV} chose both overestimated and underestimated results, leveraging the properties of overestimation and underestimation to reduce errors in the fusion process. 
In addition, the P@2 for estimator selection is 71\%, as shown in Figure \ref{fig:cmp-paradigm}, indicating that the selected overestimating and underestimating estimators are also among the top estimators.
This demonstrates the effectiveness of the top-$k$ selection strategy of \textsc{AdaNDV}.}

\noindent\textbf{Adaptivity of Base Estimators.} \textsc{AdaNDV} contains fourteen traditional base estimators and it is adaptive to add or remove base estimators. In addition, our method is available for learned estimators. To show the flexibility of base estimators in \textsc{AdaNDV}, we respectively remove the EB estimator and add the LS$_\mathrm{general}$ estimator in \textsc{AdaNDV}. The two variants are respectively named \textsc{AdaNDV}{(base}-EB{)} and \textsc{AdaNDV}{(base}+LS$_\mathrm{general}${)}, and their performance is demonstrated in Table~\ref{tab:flex}. We can derive the following conclusions based on the results: {(1)} \textsc{AdaNDV}{(base}-EB{)} and \textsc{AdaNDV}{(base}+LS$_\mathrm{general}${)} consistently outperform the individual estimators within their base estimator sets, which indicates the effectiveness of our method remains in changing the base estimators. {(2)}  Removing EB results in a consistent performance decline, indicating that estimators that are frequently selected, such as EB, play a crucial role in maintaining the overall effectiveness of the model. Adding LS$_\mathrm{general}$ will not consistently improve the performance: some metrics have decreased, while others have increased. The possible reason is that the leading estimator is relative, based on the current set of base estimators. 
Changing the base estimators will affect estimator selection.

Figure~\ref{fig:cmp-paradigm} has shown that accurately identifying the optimal base estimator from the set of base estimators is a challenging task. In the subsequent study, we explore the potential impact on our estimation performance by considering the hypothetical scenario in which this challenge is addressed.

\noindent\textbf{Upper Bound {of Estimator Selection}.} 
To further investigate the upper bound of the estimator selection component, we suppose we can exploit the optimal estimators, where P@1 is {100\%}. This variant of the model is denoted as \textsc{AdaNDV}(Hypo), and its performance is shown in Table~\ref{tab:flex}. The results can derive the following findings: {(1)} If we can accurately select the optimal estimator with the lowest overestimated and underestimated q-error, the performance of \textsc{AdaNDV} will significantly improve. It shows the promising performance upper bound of our \textsc{AdaNDV} framework.
{(2)} Compared to the Hypo-optimal estimator in Table~\ref{tab:ideal-traditional}, the performance of \textsc{AdaNDV}(Hypo) is better in most metrics except for 50\% quantile q-error. It further demonstrates the effectiveness of \textsc{AdaNDV} and indicates that our proposed complementary estimator exploitation strategy is beneficial.

The P@1 of \textsc{AdaNDV} stands at {53\%}, indicating that there is significant scope for enhancement within our proposed method. It is feasible to explore the estimator selection methods beyond the classification and ranking paradigms in the future.

\subsection{{Further Evaluation}}\label{sec:dataanalysis}
{
In this section, we show the performance of our method under more evaluation scenarios.}

\noindent\textbf{{Performance on Artificial Distributions.}} { \textsc{AdaNDV} is trained and evaluated on the data distributions from the real world, as illustrated in Section~\ref{sec:exp-settings}. However, most previous works are evaluated on artificial data distributions, so it is not clear will baseline learned methods outperform \textsc{AdaNDV} with artificial data distribution. We conduct experiments on Zipfian distribution with skew factors ($s$) 1.2, 1.5, and 2.0, consistent with previous work~\cite{li2022sampling}. We freeze the parameters of LS$_{\mathrm{scratch}}$, LS$_{\mathrm{FT}}$, and \textsc{AdaNDV} trained on the TabLib training set, and evaluate them when sampling 1\% of data from Zipfian distributions with data size of 1e5 and 1e6. The results are shown in Table~\ref{tab:zipf} and we can draw the following conclusions. No single estimator achieves optimal results under the Zipfian distributions with different skew factors and \textsc{AdaNDV} consistently beats LS$_{\mathrm{general}}$, demonstrating that our method does not fail on standard artificial distributions. Besides, LS$_{\mathrm{general}}$ is pre-trained on an artificial dataset containing 7.2$\times 10^5$ data points~\cite{ls_wu2022learning}, in which the original columns follow specific data distributions, but it processes
the worst performance in each metric. This demonstrates the effectiveness of training models on real-world data. 
}

\begin{table}[t]
    \centering
    \caption{{Q-error of learned estimators when sampling 1\% data from Zipfian distribution with skew factors ($s$) of \{1.2, 1.5, 2.0\} with column size ($N$) of 1e5 and 1e6.}}
    \label{tab:zipf}
    \begin{tabular}{c|ccc|cccccc}
\toprule
       $N$ & \multicolumn{3}{c}{1e5} & \multicolumn{3}{c}{1e6} \\
    $s$  &  1.2 & 1.5 & 2.0  &  1.2 & 1.5 & 2.0    \\
\hline
    LS$_{\mathrm{general}}$ & 5.14& 2.93& 2.47& 3.76& 5.94& 2.5 \\
    LS$_{\mathrm{scratch}}$  & \textbf{1.27}& 1.47& 1.36& 1.53& \textbf{2.67}& \textbf{1.06}  \\
    LS$_{\mathrm{FT}}$ & 1.62& 1.56& \textbf{1.28}& 1.23& 2.93& 1.15 \\
    \textsc{AdaNDV} & 1.77& \textbf{1.20}& 2.36& \textbf{1.13}& 5.86& 1.75\\

\bottomrule
    \end{tabular}
\end{table}

\begin{figure}[t]
    \centering
    \includegraphics[width=\linewidth]{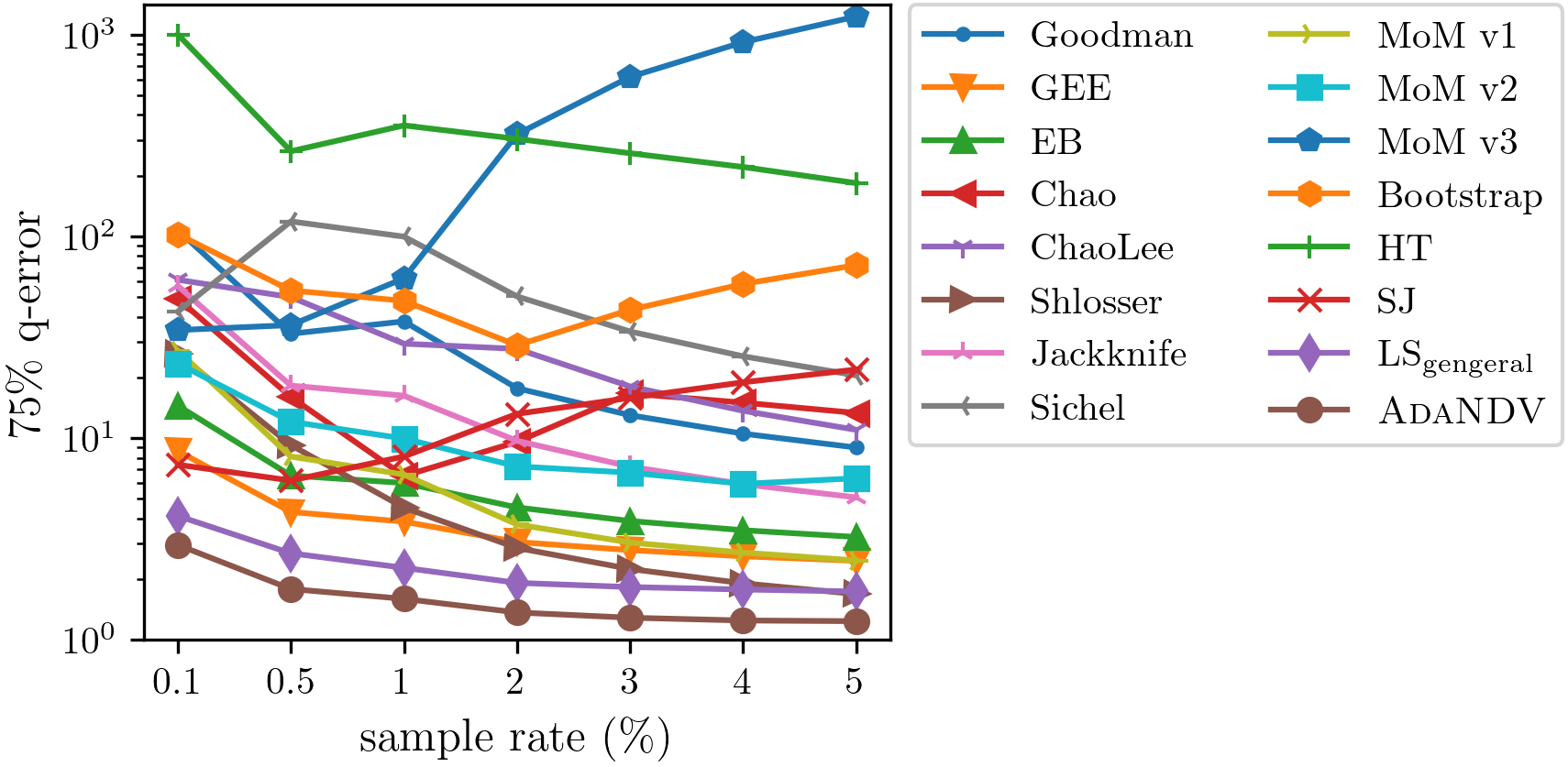}
    \caption{{Performance under different sampling rates.}}
    \label{fig:samplingrates}
\end{figure}

\noindent\textbf{{Performance under Different Sampling Rates.}} 
{
We depict the performance of 75\% quantile q-error of \textsc{AdaNDV}, base estimators, and the representative learned baseline LS$_{\mathrm{general}}$ under different sampling rates in Figure~\ref{fig:samplingrates}. \textsc{AdaNDV} shows consistent performance improvement with increasing sample rates, while some base estimators decline in performance, possibly due to practical scenarios not aligning with their assumptions. Besides, the advantages of \textsc{AdaNDV} persist across different sample rates, demonstrating that it is not sensitive to the variations of base estimators.
}

\section{Related Works}
\subsection{Sketch-based NDV Estimation}
Sketch-based NDV estimation~\cite{harmouch2017cardinality,flajolet2007hyperloglog,ertl2023ultraloglog} represents an orthogonal approach to sampling-based NDV estimation. This line of research requires scanning all the data to maintain a memory-efficient sketch for NDV estimation, which may bring an unaffordable overhead~\cite{li2022sampling}. Furthermore, real-world databases may have data access restrictions, which makes sketch-based NDV estimation not applicable in many applications. 
\subsection{Sampling-based NDV Estimation}
\textbf{Traditional NDV Estimators}. Traditional methods explore statistical techniques to summarize heuristic rules to estimate NDV and they have been studied for over seven decades in Biology~\cite{valiant2013estimating,valiant2017estimating,mmo_bunge1993estimating}, Statistics~\cite{goodman1949estimation,chao1984nonparametric}, Networks~\cite{network_cohen2019cardinality,network_nath2008synopsis}, and Databases~\cite{spark_plan_code,pg_plan_code,mysql_join}. Representative traditional estimators make different assumptions, for example, they assume infinity population size~\cite{mmo_bunge1993estimating}, certain data distribution~\cite{motwani2006distinct,mmo_bunge1993estimating}, and data skewness~\cite{hybskew_haas1995sampling,gee_charikar2000towards}. Based on the assumptions, numerous estimators have been proposed to utilize the frequency profile of sample data to build linear polinomials~\cite{goodman1949estimation,gee_charikar2000towards,error_bound}, non-linear polynomials~\cite{chao_in_db_ozsoyoglu1991estimating,chaolee,chao1984nonparametric,shlosser1981estimation,burnham1978estimation,burnham1979robust,horvitz_sarndal1992model}, and solving non-linear equations~\cite{sichel1986parameter,sichel1986word,sichel1992anatomy,bootstrap_smith1984nonparametric,mmo_bunge1993estimating,hybskew_haas1995sampling} to estimate NDV, which have been intensively discussed in Section \ref{sec:exp-settings}. In addition, some works focused on the relation between the sampling size and the errors~\cite{valiant2017estimating,wu2019chebyshev,chien2021support}.

Since representative traditional estimators are based on different heuristics, so it is difficult for them to adapt to distribution shifting.

\noindent\textbf{Learned NDV Estimators}. The introduction of ML techniques for NDV estimation has recently emerged. Wu et al.~\cite{ls_wu2022learning} are the first to leverage ML models as a Learned Statistician (LS) for NDV estimation. They improve profile maximum likelihood~\cite{apml_acharya2017unified,pml_orlitsky2004modeling,apml_pavlichin2019approximate} methods and use neural networks to take data profiles of the sampled data as inputs to estimate NDV. Li et al.~\cite{li2024learning} introduced polynomial approximation techniques~\cite{hao2019unified,wu2019chebyshev} to learn the parameters of linear polynomials of frequency profile to estimate NDV.

\subsection{Method Selection in Databases}

Selecting an optimal model from a fixed model set, as well as the ensembling multiple models, for specific database task scenarios, has emerged and garnered significant attention in recent years. Examples include identifying the proper learned cardinality estimation model for different datasets~\cite{autoce}, allocating a suitable budget for each data sampler~\cite{pengOneSizeDoes2022a}, and choosing the optimal knob tuning optimizer for each iteration~\cite{zhang2024efficient}. 
However, few studies have attempted to investigate how to select or ensemble existing NDV estimators to acquire better results. 

\section{Conclusion and future work}
In this paper, we propose \textsc{AdaNDV} to address the historically neglected selection dilemma and underexploitation issue of NDV estimators. We propose a complementary perspective of overestimated and underestimated estimators for estimation error correction. Besides, we propose fusing the estimations of the selected estimators to improve the estimation precision through a learned weighted sum, rather than directly estimating NDV. Extensive experiments on a large-scale real-life dataset exhibit the superior performance of our method.

We reveal that using existing estimators can bring promising results, however, the representation ability and the implementation paradigm of the estimator selection and fusion may limit the estimator selection precisions. In the future, we plan to develop more powerful feature extraction methods and explore more beneficial model architectures. Moreover, improving the time and space efficiency is another important direction for future work.

\bibliographystyle{ACM-Reference-Format}
\bibliography{ai4db,ref}


\begin{thebibliography}{61}


\ifx \showCODEN    \undefined \def \showCODEN     #1{\unskip}     \fi
\ifx \showDOI      \undefined \def \showDOI       #1{#1}\fi
\ifx \showISBNx    \undefined \def \showISBNx     #1{\unskip}     \fi
\ifx \showISBNxiii \undefined \def \showISBNxiii  #1{\unskip}     \fi
\ifx \showISSN     \undefined \def \showISSN      #1{\unskip}     \fi
\ifx \showLCCN     \undefined \def \showLCCN      #1{\unskip}     \fi
\ifx \shownote     \undefined \def \shownote      #1{#1}          \fi
\ifx \showarticletitle \undefined \def \showarticletitle #1{#1}   \fi
\ifx \showURL      \undefined \def \showURL       {\relax}        \fi
\providecommand\bibfield[2]{#2}
\providecommand\bibinfo[2]{#2}
\providecommand\natexlab[1]{#1}
\providecommand\showeprint[2][]{arXiv:#2}

\bibitem[\protect\citeauthoryear{??}{Cam}{2020}]%
        {Campaign}
 \bibinfo{year}{2020}\natexlab{}.
\newblock \bibinfo{title}{Campaign finance data.}
\newblock
\newblock
\urldef\tempurl%
\url{https://www.fec.gov/data/}
\showURL{%
\tempurl}


\bibitem[\protect\citeauthoryear{??}{NCV}{2020}]%
        {NCVR}
 \bibinfo{year}{2020}\natexlab{}.
\newblock \bibinfo{title}{Voter Registration Statistics.}
\newblock
\newblock
\urldef\tempurl%
\url{https://www.ncsbe.gov/results-data/voterregistration-data}
\showURL{%
\tempurl}


\bibitem[\protect\citeauthoryear{??}{box}{2024}]%
        {boxplot}
 \bibinfo{year}{2024}\natexlab{}.
\newblock \bibinfo{title}{Box plot.}
\newblock
\newblock
\urldef\tempurl%
\url{https://en.wikipedia.org/wiki/Box_plot}
\showURL{%
\tempurl}


\bibitem[\protect\citeauthoryear{??}{chi}{2024}]%
        {chi2test}
 \bibinfo{year}{2024}\natexlab{}.
\newblock \bibinfo{title}{Chi-squared test.}
\newblock
\newblock
\urldef\tempurl%
\url{https://en.wikipedia.org/wiki/Chi-squared_test}
\showURL{%
\tempurl}


\bibitem[\protect\citeauthoryear{??}{com}{2024}]%
        {commoncrawl}
 \bibinfo{year}{2024}\natexlab{}.
\newblock \bibinfo{title}{Commoncrawl.}
\newblock
\newblock
\urldef\tempurl%
\url{https://commoncrawl.org/}
\showURL{%
\tempurl}


\bibitem[\protect\citeauthoryear{??}{git}{2024}]%
        {github}
 \bibinfo{year}{2024}\natexlab{}.
\newblock \bibinfo{title}{GitHub.}
\newblock
\newblock
\urldef\tempurl%
\url{https://github.com/}
\showURL{%
\tempurl}


\bibitem[\protect\citeauthoryear{??}{ndv}{2024}]%
        {ndvlib}
 \bibinfo{year}{2024}\natexlab{}.
\newblock \bibinfo{title}{Pydistinct - Population Distinct Value Estimators.}
\newblock
\newblock
\urldef\tempurl%
\url{https://pydistinct.readthedocs.io/}
\showURL{%
\tempurl}


\bibitem[\protect\citeauthoryear{??}{mys}{2024}]%
        {mysql_join}
 \bibinfo{year}{2024}\natexlab{}.
\newblock \bibinfo{title}{Source Code of MySQL.}
\newblock
\newblock
\urldef\tempurl%
\url{https://github.com/mysql/mysql-server/blob/824e2b4064053f7daf17d7f3f84b7a3ed92e5fb4/sql/join_optimizer/cost_model.cc}
\showURL{%
\tempurl}


\bibitem[\protect\citeauthoryear{??}{pg_}{2024}]%
        {pg_plan_code}
 \bibinfo{year}{2024}\natexlab{}.
\newblock \bibinfo{title}{Source Code of PostgreSQL.}
\newblock
\newblock
\urldef\tempurl%
\url{https://github.com/postgres/postgres/blob/master/src/backend/optimizer/plan/analyzejoins.c}
\showURL{%
\tempurl}


\bibitem[\protect\citeauthoryear{??}{spa}{2024}]%
        {spark_plan_code}
 \bibinfo{year}{2024}\natexlab{}.
\newblock \bibinfo{title}{Source Code of Spark.}
\newblock
\newblock
\urldef\tempurl%
\url{https://github.com/apache/spark/blob/master/sql/catalyst/src/main/scala/org/apache/spark/sql/catalyst/plans/logical/statsEstimation/JoinEstimation.scala}
\showURL{%
\tempurl}


\bibitem[\protect\citeauthoryear{??}{tab}{2024}]%
        {tablib-v1-sample}
 \bibinfo{year}{2024}\natexlab{}.
\newblock \bibinfo{title}{tablib-v1-sample dataset.}
\newblock
\newblock
\urldef\tempurl%
\url{https://huggingface.co/datasets/approximatelabs/tablib-v1-sample}
\showURL{%
\tempurl}


\bibitem[\protect\citeauthoryear{??}{vio}{2024}]%
        {violinplot}
 \bibinfo{year}{2024}\natexlab{}.
\newblock \bibinfo{title}{Violin plot.}
\newblock
\newblock
\urldef\tempurl%
\url{https://en.wikipedia.org/wiki/Violin_plot}
\showURL{%
\tempurl}


\bibitem[\protect\citeauthoryear{Acharya, Das, Orlitsky, and Suresh}{Acharya et~al\mbox{.}}{2017}]%
        {apml_acharya2017unified}
\bibfield{author}{\bibinfo{person}{Jayadev Acharya}, \bibinfo{person}{Hirakendu Das}, \bibinfo{person}{Alon Orlitsky}, {and} \bibinfo{person}{Ananda~Theertha Suresh}.} \bibinfo{year}{2017}\natexlab{}.
\newblock \showarticletitle{A unified maximum likelihood approach for estimating symmetric properties of discrete distributions}. In \bibinfo{booktitle}{\emph{International Conference on Machine Learning}}. PMLR, \bibinfo{pages}{11--21}.
\newblock


\bibitem[\protect\citeauthoryear{Bruch, Zoghi, Bendersky, and Najork}{Bruch et~al\mbox{.}}{2019}]%
        {bruch2019revisiting}
\bibfield{author}{\bibinfo{person}{Sebastian Bruch}, \bibinfo{person}{Masrour Zoghi}, \bibinfo{person}{Michael Bendersky}, {and} \bibinfo{person}{Marc Najork}.} \bibinfo{year}{2019}\natexlab{}.
\newblock \showarticletitle{Revisiting approximate metric optimization in the age of deep neural networks}. In \bibinfo{booktitle}{\emph{Proceedings of the 42nd international ACM SIGIR conference on research and development in information retrieval}}. \bibinfo{pages}{1241--1244}.
\newblock


\bibitem[\protect\citeauthoryear{Bunge and Fitzpatrick}{Bunge and Fitzpatrick}{1993}]%
        {mmo_bunge1993estimating}
\bibfield{author}{\bibinfo{person}{John Bunge} {and} \bibinfo{person}{Michael Fitzpatrick}.} \bibinfo{year}{1993}\natexlab{}.
\newblock \showarticletitle{Estimating the number of species: a review}.
\newblock \bibinfo{journal}{\emph{Journal of the American statistical Association}} \bibinfo{volume}{88}, \bibinfo{number}{421} (\bibinfo{year}{1993}), \bibinfo{pages}{364--373}.
\newblock


\bibitem[\protect\citeauthoryear{Burnham and Overton}{Burnham and Overton}{1978}]%
        {burnham1978estimation}
\bibfield{author}{\bibinfo{person}{Kenneth~P Burnham} {and} \bibinfo{person}{Walter~Scott Overton}.} \bibinfo{year}{1978}\natexlab{}.
\newblock \showarticletitle{Estimation of the size of a closed population when capture probabilities vary among animals}.
\newblock \bibinfo{journal}{\emph{Biometrika}} \bibinfo{volume}{65}, \bibinfo{number}{3} (\bibinfo{year}{1978}), \bibinfo{pages}{625--633}.
\newblock


\bibitem[\protect\citeauthoryear{Burnham and Overton}{Burnham and Overton}{1979}]%
        {burnham1979robust}
\bibfield{author}{\bibinfo{person}{Kenneth~P Burnham} {and} \bibinfo{person}{W~Scott Overton}.} \bibinfo{year}{1979}\natexlab{}.
\newblock \showarticletitle{Robust estimation of population size when capture probabilities vary among animals}.
\newblock \bibinfo{journal}{\emph{Ecology}} \bibinfo{volume}{60}, \bibinfo{number}{5} (\bibinfo{year}{1979}), \bibinfo{pages}{927--936}.
\newblock


\bibitem[\protect\citeauthoryear{Chao}{Chao}{1984}]%
        {chao1984nonparametric}
\bibfield{author}{\bibinfo{person}{Anne Chao}.} \bibinfo{year}{1984}\natexlab{}.
\newblock \showarticletitle{Nonparametric estimation of the number of classes in a population}.
\newblock \bibinfo{journal}{\emph{Scandinavian Journal of statistics}} (\bibinfo{year}{1984}), \bibinfo{pages}{265--270}.
\newblock


\bibitem[\protect\citeauthoryear{Chao and Lee}{Chao and Lee}{1992}]%
        {chaolee}
\bibfield{author}{\bibinfo{person}{Anne Chao} {and} \bibinfo{person}{Shen-Ming Lee}.} \bibinfo{year}{1992}\natexlab{}.
\newblock \showarticletitle{Estimating the number of classes via sample coverage}.
\newblock \bibinfo{journal}{\emph{Journal of the American statistical Association}} \bibinfo{volume}{87}, \bibinfo{number}{417} (\bibinfo{year}{1992}), \bibinfo{pages}{210--217}.
\newblock


\bibitem[\protect\citeauthoryear{Charikar, Chaudhuri, Motwani, and Narasayya}{Charikar et~al\mbox{.}}{2000}]%
        {gee_charikar2000towards}
\bibfield{author}{\bibinfo{person}{Moses Charikar}, \bibinfo{person}{Surajit Chaudhuri}, \bibinfo{person}{Rajeev Motwani}, {and} \bibinfo{person}{Vivek Narasayya}.} \bibinfo{year}{2000}\natexlab{}.
\newblock \showarticletitle{Towards estimation error guarantees for distinct values}. In \bibinfo{booktitle}{\emph{Proceedings of the nineteenth ACM SIGMOD-SIGACT-SIGART symposium on Principles of database systems}}. \bibinfo{pages}{268--279}.
\newblock


\bibitem[\protect\citeauthoryear{Chaudhuri, Motwani, and Narasayya}{Chaudhuri et~al\mbox{.}}{1998}]%
        {error_bound}
\bibfield{author}{\bibinfo{person}{Surajit Chaudhuri}, \bibinfo{person}{Rajeev Motwani}, {and} \bibinfo{person}{Vivek Narasayya}.} \bibinfo{year}{1998}\natexlab{}.
\newblock \showarticletitle{Random sampling for histogram construction: How much is enough?}
\newblock \bibinfo{journal}{\emph{ACM SIGMOD Record}} \bibinfo{volume}{27}, \bibinfo{number}{2} (\bibinfo{year}{1998}), \bibinfo{pages}{436--447}.
\newblock


\bibitem[\protect\citeauthoryear{Chien, Milenkovic, and Nedich}{Chien et~al\mbox{.}}{2021}]%
        {chien2021support}
\bibfield{author}{\bibinfo{person}{Eli Chien}, \bibinfo{person}{Olgica Milenkovic}, {and} \bibinfo{person}{Angelia Nedich}.} \bibinfo{year}{2021}\natexlab{}.
\newblock \showarticletitle{Support estimation with sampling artifacts and errors}. In \bibinfo{booktitle}{\emph{2021 IEEE International Symposium on Information Theory (ISIT)}}. IEEE, \bibinfo{pages}{244--249}.
\newblock


\bibitem[\protect\citeauthoryear{Cohen and Nezri}{Cohen and Nezri}{2019}]%
        {network_cohen2019cardinality}
\bibfield{author}{\bibinfo{person}{Reuven Cohen} {and} \bibinfo{person}{Yuval Nezri}.} \bibinfo{year}{2019}\natexlab{}.
\newblock \showarticletitle{Cardinality estimation in a virtualized network device using online machine learning}.
\newblock \bibinfo{journal}{\emph{IEEE/ACM Transactions on Networking}} \bibinfo{volume}{27}, \bibinfo{number}{5} (\bibinfo{year}{2019}), \bibinfo{pages}{2098--2110}.
\newblock


\bibitem[\protect\citeauthoryear{Eggert, Huo, Biven, and Waugh}{Eggert et~al\mbox{.}}{2023}]%
        {eggert2023tablib}
\bibfield{author}{\bibinfo{person}{Gus Eggert}, \bibinfo{person}{Kevin Huo}, \bibinfo{person}{Mike Biven}, {and} \bibinfo{person}{Justin Waugh}.} \bibinfo{year}{2023}\natexlab{}.
\newblock \bibinfo{title}{TabLib: A Dataset of 627M Tables with Context}.
\newblock
\newblock
\showeprint[arxiv]{2310.07875}~[cs.CL]


\bibitem[\protect\citeauthoryear{Ertl}{Ertl}{2024}]%
        {ertl2023ultraloglog}
\bibfield{author}{\bibinfo{person}{Otmar Ertl}.} \bibinfo{year}{2024}\natexlab{}.
\newblock \showarticletitle{UltraLogLog: A Practical and More Space-Efficient Alternative to HyperLogLog for Approximate Distinct Counting}.
\newblock \bibinfo{journal}{\emph{Proceedings of the VLDB Endowment}} \bibinfo{volume}{17}, \bibinfo{number}{7} (\bibinfo{year}{2024}), \bibinfo{pages}{1655--1668}.
\newblock


\bibitem[\protect\citeauthoryear{Flajolet, Fusy, Gandouet, and Meunier}{Flajolet et~al\mbox{.}}{2007}]%
        {flajolet2007hyperloglog}
\bibfield{author}{\bibinfo{person}{Philippe Flajolet}, \bibinfo{person}{{\'E}ric Fusy}, \bibinfo{person}{Olivier Gandouet}, {and} \bibinfo{person}{Fr{\'e}d{\'e}ric Meunier}.} \bibinfo{year}{2007}\natexlab{}.
\newblock \showarticletitle{Hyperloglog: the analysis of a near-optimal cardinality estimation algorithm}.
\newblock \bibinfo{journal}{\emph{Discrete mathematics \& theoretical computer science}} \bibinfo{number}{Proceedings} (\bibinfo{year}{2007}).
\newblock


\bibitem[\protect\citeauthoryear{Goodman}{Goodman}{1949}]%
        {goodman1949estimation}
\bibfield{author}{\bibinfo{person}{Leo~A Goodman}.} \bibinfo{year}{1949}\natexlab{}.
\newblock \showarticletitle{On the estimation of the number of classes in a population}.
\newblock \bibinfo{journal}{\emph{The Annals of Mathematical Statistics}} \bibinfo{volume}{20}, \bibinfo{number}{4} (\bibinfo{year}{1949}), \bibinfo{pages}{572--579}.
\newblock


\bibitem[\protect\citeauthoryear{Haas, Naughton, Seshadri, and Stokes}{Haas et~al\mbox{.}}{1995}]%
        {hybskew_haas1995sampling}
\bibfield{author}{\bibinfo{person}{Peter~J Haas}, \bibinfo{person}{Jeffrey~F Naughton}, \bibinfo{person}{S Seshadri}, {and} \bibinfo{person}{Lynne Stokes}.} \bibinfo{year}{1995}\natexlab{}.
\newblock \showarticletitle{Sampling-based estimation of the number of distinct values of an attribute}. In \bibinfo{booktitle}{\emph{VLDB}}, Vol.~\bibinfo{volume}{95}. \bibinfo{pages}{311--322}.
\newblock


\bibitem[\protect\citeauthoryear{Han, Wang, Chen, Dong, Chen, Yu, Yang, and Qian}{Han et~al\mbox{.}}{2024}]%
        {han2024bytecard}
\bibfield{author}{\bibinfo{person}{Yuxing Han}, \bibinfo{person}{Haoyu Wang}, \bibinfo{person}{Lixiang Chen}, \bibinfo{person}{Yifeng Dong}, \bibinfo{person}{Xing Chen}, \bibinfo{person}{Benquan Yu}, \bibinfo{person}{Chengcheng Yang}, {and} \bibinfo{person}{Weining Qian}.} \bibinfo{year}{2024}\natexlab{}.
\newblock \showarticletitle{ByteCard: Enhancing Data Warehousing with Learned Cardinality Estimation}.
\newblock \bibinfo{journal}{\emph{arXiv preprint arXiv:2403.16110}} (\bibinfo{year}{2024}).
\newblock


\bibitem[\protect\citeauthoryear{Hao and Orlitsky}{Hao and Orlitsky}{2019}]%
        {hao2019unified}
\bibfield{author}{\bibinfo{person}{Yi Hao} {and} \bibinfo{person}{Alon Orlitsky}.} \bibinfo{year}{2019}\natexlab{}.
\newblock \showarticletitle{Unified sample-optimal property estimation in near-linear time}.
\newblock \bibinfo{journal}{\emph{Advances in Neural Information Processing Systems}}  \bibinfo{volume}{32} (\bibinfo{year}{2019}).
\newblock


\bibitem[\protect\citeauthoryear{Harmouch and Naumann}{Harmouch and Naumann}{2017}]%
        {harmouch2017cardinality}
\bibfield{author}{\bibinfo{person}{Hazar Harmouch} {and} \bibinfo{person}{Felix Naumann}.} \bibinfo{year}{2017}\natexlab{}.
\newblock \showarticletitle{Cardinality estimation: An experimental survey}.
\newblock \bibinfo{journal}{\emph{Proceedings of the VLDB Endowment}} \bibinfo{volume}{11}, \bibinfo{number}{4} (\bibinfo{year}{2017}), \bibinfo{pages}{499--512}.
\newblock


\bibitem[\protect\citeauthoryear{Hulsebos, Demiralp, and Groth}{Hulsebos et~al\mbox{.}}{2023}]%
        {hulsebos2023gittables}
\bibfield{author}{\bibinfo{person}{Madelon Hulsebos}, \bibinfo{person}{{\c{C}}agatay Demiralp}, {and} \bibinfo{person}{Paul Groth}.} \bibinfo{year}{2023}\natexlab{}.
\newblock \showarticletitle{Gittables: A large-scale corpus of relational tables}.
\newblock \bibinfo{journal}{\emph{Proceedings of the ACM on Management of Data}} \bibinfo{volume}{1}, \bibinfo{number}{1} (\bibinfo{year}{2023}), \bibinfo{pages}{1--17}.
\newblock


\bibitem[\protect\citeauthoryear{Kingma and Ba}{Kingma and Ba}{2015}]%
        {adam_kingma2014adam}
\bibfield{author}{\bibinfo{person}{Diederik~P. Kingma} {and} \bibinfo{person}{Jimmy Ba}.} \bibinfo{year}{2015}\natexlab{}.
\newblock \showarticletitle{Adam: {A} Method for Stochastic Optimization}. In \bibinfo{booktitle}{\emph{3rd International Conference on Learning Representations, {ICLR} 2015, San Diego, CA, USA, May 7-9, 2015, Conference Track Proceedings}}.
\newblock


\bibitem[\protect\citeauthoryear{Li, Lei, Wang, Wei, and Ding}{Li et~al\mbox{.}}{2024}]%
        {li2024learning}
\bibfield{author}{\bibinfo{person}{Jiajun Li}, \bibinfo{person}{Runlin Lei}, \bibinfo{person}{Sibo Wang}, \bibinfo{person}{Zhewei Wei}, {and} \bibinfo{person}{Bolin Ding}.} \bibinfo{year}{2024}\natexlab{}.
\newblock \showarticletitle{Learning-based Property Estimation with Polynomials}.
\newblock \bibinfo{journal}{\emph{Proceedings of the ACM on Management of Data}} \bibinfo{volume}{2}, \bibinfo{number}{3} (\bibinfo{year}{2024}), \bibinfo{pages}{1--27}.
\newblock


\bibitem[\protect\citeauthoryear{Li, Wei, Ding, Dai, Lu, and Zhou}{Li et~al\mbox{.}}{2022}]%
        {li2022sampling}
\bibfield{author}{\bibinfo{person}{Jiajun Li}, \bibinfo{person}{Zhewei Wei}, \bibinfo{person}{Bolin Ding}, \bibinfo{person}{Xiening Dai}, \bibinfo{person}{Lu Lu}, {and} \bibinfo{person}{Jingren Zhou}.} \bibinfo{year}{2022}\natexlab{}.
\newblock \showarticletitle{Sampling-based estimation of the number of distinct values in distributed environment}. In \bibinfo{booktitle}{\emph{Proceedings of the 28th ACM SIGKDD Conference on Knowledge Discovery and Data Mining}}. \bibinfo{pages}{893--903}.
\newblock


\bibitem[\protect\citeauthoryear{Li, Wei, Zhu, Ding, Zhou, and Lu}{Li et~al\mbox{.}}{2023}]%
        {li2023alece}
\bibfield{author}{\bibinfo{person}{Pengfei Li}, \bibinfo{person}{Wenqing Wei}, \bibinfo{person}{Rong Zhu}, \bibinfo{person}{Bolin Ding}, \bibinfo{person}{Jingren Zhou}, {and} \bibinfo{person}{Hua Lu}.} \bibinfo{year}{2023}\natexlab{}.
\newblock \showarticletitle{ALECE: An Attention-based Learned Cardinality Estimator for SPJ Queries on Dynamic Workloads}.
\newblock \bibinfo{journal}{\emph{Proceedings of the VLDB Endowment}} \bibinfo{volume}{17}, \bibinfo{number}{2} (\bibinfo{year}{2023}), \bibinfo{pages}{197--210}.
\newblock


\bibitem[\protect\citeauthoryear{Liu et~al\mbox{.}}{Liu et~al\mbox{.}}{2009}]%
        {liu2009learning}
\bibfield{author}{\bibinfo{person}{Tie-Yan Liu} {et~al\mbox{.}}} \bibinfo{year}{2009}\natexlab{}.
\newblock \showarticletitle{Learning to rank for information retrieval}.
\newblock \bibinfo{journal}{\emph{Foundations and Trends{\textregistered} in Information Retrieval}} \bibinfo{volume}{3}, \bibinfo{number}{3} (\bibinfo{year}{2009}), \bibinfo{pages}{225--331}.
\newblock


\bibitem[\protect\citeauthoryear{Moerkotte, Neumann, and Steidl}{Moerkotte et~al\mbox{.}}{2009}]%
        {q_error_moerkotte2009preventing}
\bibfield{author}{\bibinfo{person}{Guido Moerkotte}, \bibinfo{person}{Thomas Neumann}, {and} \bibinfo{person}{Gabriele Steidl}.} \bibinfo{year}{2009}\natexlab{}.
\newblock \showarticletitle{Preventing bad plans by bounding the impact of cardinality estimation errors}.
\newblock \bibinfo{journal}{\emph{Proceedings of the VLDB Endowment}} \bibinfo{volume}{2}, \bibinfo{number}{1} (\bibinfo{year}{2009}), \bibinfo{pages}{982--993}.
\newblock


\bibitem[\protect\citeauthoryear{Motwani and Vassilvitskii}{Motwani and Vassilvitskii}{2006}]%
        {motwani2006distinct}
\bibfield{author}{\bibinfo{person}{Rajeev Motwani} {and} \bibinfo{person}{Sergei Vassilvitskii}.} \bibinfo{year}{2006}\natexlab{}.
\newblock \showarticletitle{Distinct values estimators for power law distributions}. In \bibinfo{booktitle}{\emph{2006 Proceedings of the Third Workshop on Analytic Algorithmics and Combinatorics (ANALCO)}}. SIAM, \bibinfo{pages}{230--237}.
\newblock


\bibitem[\protect\citeauthoryear{Nath, Gibbons, Seshan, and Anderson}{Nath et~al\mbox{.}}{2008}]%
        {network_nath2008synopsis}
\bibfield{author}{\bibinfo{person}{Suman Nath}, \bibinfo{person}{Phillip~B Gibbons}, \bibinfo{person}{Srinivasan Seshan}, {and} \bibinfo{person}{Zachary Anderson}.} \bibinfo{year}{2008}\natexlab{}.
\newblock \showarticletitle{Synopsis diffusion for robust aggregation in sensor networks}.
\newblock \bibinfo{journal}{\emph{ACM Transactions on Sensor Networks (TOSN)}} \bibinfo{volume}{4}, \bibinfo{number}{2} (\bibinfo{year}{2008}), \bibinfo{pages}{1--40}.
\newblock


\bibitem[\protect\citeauthoryear{Orlitsky, Santhanam, Viswanathan, and Zhang}{Orlitsky et~al\mbox{.}}{2004}]%
        {pml_orlitsky2004modeling}
\bibfield{author}{\bibinfo{person}{Alon Orlitsky}, \bibinfo{person}{Narayana~P Santhanam}, \bibinfo{person}{Krishnamurthy Viswanathan}, {and} \bibinfo{person}{Junan Zhang}.} \bibinfo{year}{2004}\natexlab{}.
\newblock \showarticletitle{On modeling profiles instead of values}. In \bibinfo{booktitle}{\emph{Proceedings of the 20th conference on Uncertainty in artificial intelligence}}. \bibinfo{pages}{426--435}.
\newblock


\bibitem[\protect\citeauthoryear{Ozsoyoglu, Du, Tjahjana, Hou, and Rowland}{Ozsoyoglu et~al\mbox{.}}{1991}]%
        {chao_in_db_ozsoyoglu1991estimating}
\bibfield{author}{\bibinfo{person}{Gultekin Ozsoyoglu}, \bibinfo{person}{Kaizheng Du}, \bibinfo{person}{A Tjahjana}, \bibinfo{person}{W-C Hou}, {and} \bibinfo{person}{DY Rowland}.} \bibinfo{year}{1991}\natexlab{}.
\newblock \showarticletitle{On estimating COUNT, SUM, and AVERAGE relational algebra queries}. In \bibinfo{booktitle}{\emph{Database and Expert Systems Applications: Proceedings of the International Conference in Berlin, Federal Republic of Germany, 1991}}. Springer, \bibinfo{pages}{406--412}.
\newblock


\bibitem[\protect\citeauthoryear{O’Neil, O’Neil, Chen, and Revilak}{O’Neil et~al\mbox{.}}{2009}]%
        {ssb_o2009star}
\bibfield{author}{\bibinfo{person}{Patrick O’Neil}, \bibinfo{person}{Elizabeth O’Neil}, \bibinfo{person}{Xuedong Chen}, {and} \bibinfo{person}{Stephen Revilak}.} \bibinfo{year}{2009}\natexlab{}.
\newblock \showarticletitle{The star schema benchmark and augmented fact table indexing}. In \bibinfo{booktitle}{\emph{Performance Evaluation and Benchmarking: First TPC Technology Conference, TPCTC 2009, Lyon, France, August 24-28, 2009, Revised Selected Papers 1}}. Springer, \bibinfo{pages}{237--252}.
\newblock


\bibitem[\protect\citeauthoryear{Pasumarthi, Bruch, Wang, Li, Bendersky, Najork, Pfeifer, Golbandi, Anil, and Wolf}{Pasumarthi et~al\mbox{.}}{2019}]%
        {TensorflowRankingKDD2019}
\bibfield{author}{\bibinfo{person}{Rama~Kumar Pasumarthi}, \bibinfo{person}{Sebastian Bruch}, \bibinfo{person}{Xuanhui Wang}, \bibinfo{person}{Cheng Li}, \bibinfo{person}{Michael Bendersky}, \bibinfo{person}{Marc Najork}, \bibinfo{person}{Jan Pfeifer}, \bibinfo{person}{Nadav Golbandi}, \bibinfo{person}{Rohan Anil}, {and} \bibinfo{person}{Stephan Wolf}.} \bibinfo{year}{2019}\natexlab{}.
\newblock \showarticletitle{TF-Ranking: Scalable TensorFlow Library for Learning-to-Rank}. In \bibinfo{booktitle}{\emph{Proceedings of the 25th ACM SIGKDD International Conference on Knowledge Discovery and Data Mining}} (Anchorage, AK). \bibinfo{pages}{2970--2978}.
\newblock


\bibitem[\protect\citeauthoryear{Pavlichin, Jiao, and Weissman}{Pavlichin et~al\mbox{.}}{2019}]%
        {apml_pavlichin2019approximate}
\bibfield{author}{\bibinfo{person}{Dmitri~S Pavlichin}, \bibinfo{person}{Jiantao Jiao}, {and} \bibinfo{person}{Tsachy Weissman}.} \bibinfo{year}{2019}\natexlab{}.
\newblock \showarticletitle{Approximate profile maximum likelihood}.
\newblock \bibinfo{journal}{\emph{Journal of Machine Learning Research}} \bibinfo{volume}{20}, \bibinfo{number}{122} (\bibinfo{year}{2019}), \bibinfo{pages}{1--55}.
\newblock


\bibitem[\protect\citeauthoryear{Peng, Ding, Wang, Zeng, and Zhou}{Peng et~al\mbox{.}}{2022}]%
        {pengOneSizeDoes2022a}
\bibfield{author}{\bibinfo{person}{Jinglin Peng}, \bibinfo{person}{Bolin Ding}, \bibinfo{person}{Jiannan Wang}, \bibinfo{person}{Kai Zeng}, {and} \bibinfo{person}{Jingren Zhou}.} \bibinfo{year}{2022}\natexlab{}.
\newblock \showarticletitle{One {{Size Does Not Fit All}}: {{A Bandit-Based Sampler Combination Framework}} with {{Theoretical Guarantees}}}. In \bibinfo{booktitle}{\emph{Proceedings of the 2022 {{International Conference}} on {{Management}} of {{Data}}}}. \bibinfo{publisher}{ACM}, \bibinfo{address}{Philadelphia PA USA}, \bibinfo{pages}{531--544}.
\newblock
\showISBNx{978-1-4503-9249-5}
\urldef\tempurl%
\url{https://doi.org/10.1145/3514221.3517900}
\showDOI{\tempurl}


\bibitem[\protect\citeauthoryear{S{\"a}rndal, Swensson, and Wretman}{S{\"a}rndal et~al\mbox{.}}{1992}]%
        {horvitz_sarndal1992model}
\bibfield{author}{\bibinfo{person}{Carl-Erik S{\"a}rndal}, \bibinfo{person}{Bengt Swensson}, {and} \bibinfo{person}{Jan Wretman}.} \bibinfo{year}{1992}\natexlab{}.
\newblock \showarticletitle{Model Assisted Survey Sampling}.
\newblock \bibinfo{journal}{\emph{Springer Series in Statistics}} (\bibinfo{year}{1992}).
\newblock


\bibitem[\protect\citeauthoryear{Shlosser}{Shlosser}{1981}]%
        {shlosser1981estimation}
\bibfield{author}{\bibinfo{person}{A Shlosser}.} \bibinfo{year}{1981}\natexlab{}.
\newblock \showarticletitle{On estimation of the size of the dictionary of a long text on the basis of a sample}.
\newblock \bibinfo{journal}{\emph{Engineering Cybernetics}} \bibinfo{volume}{19}, \bibinfo{number}{1} (\bibinfo{year}{1981}), \bibinfo{pages}{97--102}.
\newblock


\bibitem[\protect\citeauthoryear{Sichel}{Sichel}{1986a}]%
        {sichel1986parameter}
\bibfield{author}{\bibinfo{person}{HS Sichel}.} \bibinfo{year}{1986}\natexlab{a}.
\newblock \showarticletitle{Parameter estimation for a word frequency distribution based on occupancy theory}.
\newblock \bibinfo{journal}{\emph{Communications in Statistics-Theory and Methods}} \bibinfo{volume}{15}, \bibinfo{number}{3} (\bibinfo{year}{1986}), \bibinfo{pages}{935--949}.
\newblock


\bibitem[\protect\citeauthoryear{Sichel}{Sichel}{1986b}]%
        {sichel1986word}
\bibfield{author}{\bibinfo{person}{Herbert~S Sichel}.} \bibinfo{year}{1986}\natexlab{b}.
\newblock \showarticletitle{Word frequency distributions and type-token characteristics}.
\newblock \bibinfo{journal}{\emph{Math. Scientist}}  \bibinfo{volume}{11} (\bibinfo{year}{1986}), \bibinfo{pages}{45--72}.
\newblock


\bibitem[\protect\citeauthoryear{Sichel}{Sichel}{1992}]%
        {sichel1992anatomy}
\bibfield{author}{\bibinfo{person}{HERBERT~S Sichel}.} \bibinfo{year}{1992}\natexlab{}.
\newblock \showarticletitle{Anatomy of the generalized inverse Gaussian-Poisson distribution with special applications to bibliometric studies}.
\newblock \bibinfo{journal}{\emph{Information Processing \& Management}} \bibinfo{volume}{28}, \bibinfo{number}{1} (\bibinfo{year}{1992}), \bibinfo{pages}{5--17}.
\newblock


\bibitem[\protect\citeauthoryear{Smith and van Belle}{Smith and van Belle}{1984}]%
        {bootstrap_smith1984nonparametric}
\bibfield{author}{\bibinfo{person}{Eric~P Smith} {and} \bibinfo{person}{Gerald van Belle}.} \bibinfo{year}{1984}\natexlab{}.
\newblock \showarticletitle{Nonparametric estimation of species richness}.
\newblock \bibinfo{journal}{\emph{Biometrics}} (\bibinfo{year}{1984}), \bibinfo{pages}{119--129}.
\newblock


\bibitem[\protect\citeauthoryear{Valiant and Valiant}{Valiant and Valiant}{2017}]%
        {valiant2017estimating}
\bibfield{author}{\bibinfo{person}{Gregory Valiant} {and} \bibinfo{person}{Paul Valiant}.} \bibinfo{year}{2017}\natexlab{}.
\newblock \showarticletitle{Estimating the unseen: improved estimators for entropy and other properties}.
\newblock \bibinfo{journal}{\emph{Journal of the ACM (JACM)}} \bibinfo{volume}{64}, \bibinfo{number}{6} (\bibinfo{year}{2017}), \bibinfo{pages}{1--41}.
\newblock


\bibitem[\protect\citeauthoryear{Valiant and Valiant}{Valiant and Valiant}{2013}]%
        {valiant2013estimating}
\bibfield{author}{\bibinfo{person}{Paul Valiant} {and} \bibinfo{person}{Gregory Valiant}.} \bibinfo{year}{2013}\natexlab{}.
\newblock \showarticletitle{Estimating the Unseen: Improved Estimators for Entropy and other Properties}.
\newblock \bibinfo{journal}{\emph{Advances in Neural Information Processing Systems}}  \bibinfo{volume}{26} (\bibinfo{year}{2013}).
\newblock


\bibitem[\protect\citeauthoryear{Wang, Li, Golbandi, Bendersky, and Najork}{Wang et~al\mbox{.}}{2018}]%
        {wang2018lambdaloss}
\bibfield{author}{\bibinfo{person}{Xuanhui Wang}, \bibinfo{person}{Cheng Li}, \bibinfo{person}{Nadav Golbandi}, \bibinfo{person}{Michael Bendersky}, {and} \bibinfo{person}{Marc Najork}.} \bibinfo{year}{2018}\natexlab{}.
\newblock \showarticletitle{The lambdaloss framework for ranking metric optimization}. In \bibinfo{booktitle}{\emph{Proceedings of the 27th ACM international conference on information and knowledge management}}. \bibinfo{pages}{1313--1322}.
\newblock


\bibitem[\protect\citeauthoryear{Wang, Wang, Li, He, and Liu}{Wang et~al\mbox{.}}{2013}]%
        {wang2013theoretical}
\bibfield{author}{\bibinfo{person}{Yining Wang}, \bibinfo{person}{Liwei Wang}, \bibinfo{person}{Yuanzhi Li}, \bibinfo{person}{Di He}, {and} \bibinfo{person}{Tie-Yan Liu}.} \bibinfo{year}{2013}\natexlab{}.
\newblock \showarticletitle{A theoretical analysis of NDCG type ranking measures}. In \bibinfo{booktitle}{\emph{Conference on learning theory}}. PMLR, \bibinfo{pages}{25--54}.
\newblock


\bibitem[\protect\citeauthoryear{Wu, Ding, Chu, Wei, Dai, Guan, and Zhou}{Wu et~al\mbox{.}}{2021}]%
        {ls_wu2022learning}
\bibfield{author}{\bibinfo{person}{Renzhi Wu}, \bibinfo{person}{Bolin Ding}, \bibinfo{person}{Xu Chu}, \bibinfo{person}{Zhewei Wei}, \bibinfo{person}{Xiening Dai}, \bibinfo{person}{Tao Guan}, {and} \bibinfo{person}{Jingren Zhou}.} \bibinfo{year}{2021}\natexlab{}.
\newblock \showarticletitle{Learning to Be a Statistician: Learned Estimator for Number of Distinct Values}.
\newblock \bibinfo{journal}{\emph{Proc. VLDB Endow.}} \bibinfo{volume}{15}, \bibinfo{number}{2} (\bibinfo{date}{oct} \bibinfo{year}{2021}), \bibinfo{pages}{272–284}.
\newblock
\showISSN{2150-8097}
\urldef\tempurl%
\url{https://doi.org/10.14778/3489496.3489508}
\showDOI{\tempurl}


\bibitem[\protect\citeauthoryear{Wu and Yang}{Wu and Yang}{2019}]%
        {wu2019chebyshev}
\bibfield{author}{\bibinfo{person}{Yihong Wu} {and} \bibinfo{person}{Pengkun Yang}.} \bibinfo{year}{2019}\natexlab{}.
\newblock \showarticletitle{Chebyshev polynomials, moment matching, and optimal estimation of the unseen}.
\newblock \bibinfo{journal}{\emph{The Annals of Statistics}} \bibinfo{volume}{47}, \bibinfo{number}{2} (\bibinfo{year}{2019}), \bibinfo{pages}{857--883}.
\newblock


\bibitem[\protect\citeauthoryear{Xia, Liu, Wang, Zhang, and Li}{Xia et~al\mbox{.}}{2008}]%
        {listmle_xia2008listwise}
\bibfield{author}{\bibinfo{person}{Fen Xia}, \bibinfo{person}{Tie-Yan Liu}, \bibinfo{person}{Jue Wang}, \bibinfo{person}{Wensheng Zhang}, {and} \bibinfo{person}{Hang Li}.} \bibinfo{year}{2008}\natexlab{}.
\newblock \showarticletitle{Listwise approach to learning to rank: theory and algorithm}. In \bibinfo{booktitle}{\emph{Proceedings of the 25th international conference on Machine learning}}. \bibinfo{pages}{1192--1199}.
\newblock


\bibitem[\protect\citeauthoryear{Zhang, Zhang, Li, and Chai}{Zhang et~al\mbox{.}}{2023b}]%
        {autoce}
\bibfield{author}{\bibinfo{person}{Jintao Zhang}, \bibinfo{person}{Chao Zhang}, \bibinfo{person}{Guoliang Li}, {and} \bibinfo{person}{Chengliang Chai}.} \bibinfo{year}{2023}\natexlab{b}.
\newblock \showarticletitle{{{AutoCE}}: {{An Accurate}} and {{Efficient Model Advisor}} for {{Learned Cardinality Estimation}}}. In \bibinfo{booktitle}{\emph{2023 {{IEEE}} 39th {{International Conference}} on {{Data Engineering}} ({{ICDE}})}}. \bibinfo{publisher}{IEEE}, \bibinfo{address}{Anaheim, CA, USA}, \bibinfo{pages}{2621--2633}.
\newblock
\showISBNx{9798350322279}
\urldef\tempurl%
\url{https://doi.org/10.1109/ICDE55515.2023.00201}
\showDOI{\tempurl}


\bibitem[\protect\citeauthoryear{Zhang, Wu, Li, Tang, Tan, Li, and Cui}{Zhang et~al\mbox{.}}{2023a}]%
        {zhang2024efficient}
\bibfield{author}{\bibinfo{person}{Xinyi Zhang}, \bibinfo{person}{Hong Wu}, \bibinfo{person}{Yang Li}, \bibinfo{person}{Zhengju Tang}, \bibinfo{person}{Jian Tan}, \bibinfo{person}{Feifei Li}, {and} \bibinfo{person}{Bin Cui}.} \bibinfo{year}{2023}\natexlab{a}.
\newblock \showarticletitle{An Efficient Transfer Learning Based Configuration Adviser for Database Tuning}.
\newblock \bibinfo{journal}{\emph{Proceedings of the VLDB Endowment}} \bibinfo{volume}{17}, \bibinfo{number}{3} (\bibinfo{year}{2023}), \bibinfo{pages}{539--552}.
\newblock


\end{thebibliography}

\end{document}